# The 1908 Tunguska event and forestfalls

Andrei Ol'khovatov

https://orcid.org/0000-0002-6043-9205

Independent researcher
(Retired physicist)

Russia, Moscow
email: olkhov@mail.ru

**Dedicated to the blessed memory of my grandmother ( Tuzlukova Anna Ivanovna ) and my mother ( Ol'khovatova Olga Leonidovna )**

**Abstract.** The 1908 Tunguska event is used to be associated with a forestfall named after its first scientific researcher – Leonid Kulik. However association of the Kulikovskii forestfall with the events in the morning of June 30, 1908 is based only on Evenks accounts. Initially, Kulik assumed the impact site was about 60 km east of this forestfall (unknown to him at that time). Then he received additional information about the existence of this large forestfall, and moved the search point.

Later, some other forestfalls were reported, albeit of a much smaller area. Dates some of them can't be pinpointed accurately, i.e. only "about June 30", and so on. However Evenks reported about several small forestfalls in the region which according to them occurred on the same morning as the Kulikovskii one. But as they were much smaller then a little attention was paid to them (with the exception of the so-called Chuvar forestfall). This paper is devoted to consideration of these forestfalls (including some peculiarities of the Kulikovskii one). The paper focuses on facts (and not on interpretations), which any interpretation of the 1908 Tunguska event should explain.

## 1. Introduction

In the morning of June 30, 1908, thunderous sounds were heard by population north and northwest of Lake Baikal in Central Siberia, In some places also the ground trembled. Reports of a flying glowing body came from various points of the region. Soon a newspaper story appeared about a fall of a large meteorite near the town of Kansk, but then it was recognized as wrong.

First reports in mass-media about damage of the forest occurred in 1908 in "hot pursuit". For example, there was an article "A meteor, a lightning or an earthquake" in the July 15, 1908 (in the Julian calendar) issue of the newspaper "Golos Tomska" which included the following text (translated by A. Ol'khovatov):

**"For some time, rumors were spreading that the aerolite fell near the village of Dalaya and as if many had seen how it flew, and that during the fall, this aerolite hit a tree - a thick pine, which it smashed and there were a lot of such stories."**

However info about large forestfalls circulated only on the level of rumors or second-hand accounts at best. Soon the event was almost forgotten. It was L. A. Kulik who was the first to raise the question in mass-media and in scientific publications in the early 1920s. Initially, Kulik assumed the meteorite impact site in the area of the river Ognia - the left upper tributary of the Vanovara (Vanovarka) river basing on reports of some destruction in this area. Later he received additional information about the existence of a large forestfall in another area and moved the search point. In 1927 he reached the large forestfall which later was named after him - Kulikovskii. In general the Kulikovskii forestfall is of radial character (i.e. tress fell outward of a place called the epicenter). The position of the epicenter was established is 1960s by a group of the Soviet Tunguska researchers who called their group as KSE (Kompleksnaya Samodeyatel'naya Ekspeditsiya). Its informal leader from the mid-1960s until his death in 2001 was Nikolai Vladimirovich Vasil'ev.

The epicenter's position is about 60.9° N, and 101.9° E. In this paper the distance will be often measured from the epicenter of the Kulikovskii forestfall (and azimuth is measured from north clockwise).

However Evenks reported about several other forestfalls in the region which according to them occurred on that day. Also there were reports (by Evenks and local residents) pointing to several more forestfalls, which dates can be pinpointed less accurately, i.e. "about June 30", and so on. But as they were much smaller than the Kulikovskii one, then a little attention was paid to them. This paper is devoted to consideration of these forestfalls. But let's start with the Kulikovskii forestfall.

## 2. The Kulikovskii forestfall

In the early 1920s L.A. Kulik investigated the case of the reported "meteorite fall", and found out that the initial newspaper story of the fall near Kansk was false.

Also he got some info from various people which directed him to another place.

The chain of events which led Kulik to the place north of Vanavara can be seen from his 1927 article, which was rather well translated in English in 1935 (Wiens and La Paz, 1935). Here are some citations from the translation:

> "Thus, a member of the management of the Consumers Association of Kezhma on the Angara, I. K. Vologzhin, reported the following to the author in the city of Krasnoyarsk on the 21st of November, 1921: " (...) In the winter the inhabitants of the village Kezhma, who traded around the Podkamennaya Tunguska (Khatanga), reported that the Tunguse Ivan Ilyich Ilyushenok related that at the time of this occurrence, in the locality where they were wandering between the Podkamennaya Tunguska and the Nishnaya [lower] Tunguska, a strip of forest was uprooted by the pressure of the air and several of his reindeer were killed. People suppose that the place of fall is in the region between the Podkamennaya Tunguska and the Nishnaya Tunguska."
> 
>   I. I. Pokrovsky (city of Yennisseisk [Yeniseisk], former financial inspector) corroborates the rumor about the breaking down of the forest by a wind blast; in a letter to the author under date of April 21, 1922, he writes: "According to the testimony of the Tunguses, the falling of the meteorite was accompanied by an unusual atmospheric perturbation which caused a terrible destruction of the woods over a large area."
> 
>   More concrete data about the destruction of the forest by a wind blast are furnished by an inhabitant of the village Krasnoyarovo, of the district of Kirensk, of the government of Irkutsk, V. M. Arbatsky, in answer to a questionnaire of the Meteorite Expedition. After giving a description of the phenomenon, containing nothing new, he adds: "In the region reached by going from the small river Tunguska along the little river Ayan for about 15 versts, I noticed a broad strip of completely uprooted forest, extending along the road for about one verst. As for the more remote regions [i.e. farther away from the road] I have no knowledge." "

Unfortunately Kulik did not mention that according to Arbatsky the thunderous sounds were heard for about an hour and a half, and they started at 10 am. We will return to the Arbatsky report later.

I. V. Kolmakov (co-operator in the village Panovskoye on the Angara) wrote to Kulik on February 10, 1922 the following (Wiens and La Paz, 1935):

"

"(...) Moreover, I, personally, talked with a Tunguse about this [event]. The latter related the following: 'Along that same river Tunguska in the region of the junction with the Chamba over a distance of 350 versts from Panovskoye, during the time of this thunder, about a thousand of the Tunguses' reindeer were killed and the remainder of the reindeer badly injured, and also the natives themselves suffered from the heavy shake [concussion], and, moreover, in a region of radius approximately 70 versts, all the forest was destroyed and right there by the shock a spring was opened in the earth which disappeared after several days, but the place of outflow of the water was not examined by the natives. All that which is contained herein I affirm under oath."
"

Early Kulik thought (on the basis of info collected in the Kansky district, and subsequently, in Tomsk and other places) that the meteorire fell in the area of the river Ognia - the left upper tributary of the Vanovara (Vanovarka) river (Kulik,1922). It will be considered with more details below.

In 1924 Kulik received a letter dated February 2, 1924, from the geologist and co-worker in the Krasnoyarsk Museum, A. N. Sobolev. Here is what Kulik wrote about it (Wiens and La Paz, 1935):

"
His account gives a vivid picture of this mighty phenomenon and completely agrees with the data of I. M. Suslov: "A certain N. N. Kartashev working in the summer on the Podkamennaya Tunguska reports: 'According to the story of a Tunguse Ilya Potapovich [no surname], who lives on the river Tetera in the upper regions of the Podkamennaya Tunguska, long ago (about 15 years), there lived on the river Chamba his brother [who is now an old Tunguse and who hardly speaks Russian, and who now lives on the Tetera with Ilya Potapovich and whom N. N. Kartashev saw]. Ilya Potapovich reported that there [on the river Chamba], once upon a time, occurred some sort of terrible explosion with noise and wind. The power of the explosion was so great that, on the river, for many versts along both sides, the forest was broken down in one direction. The reindeer skin tent of his brother was thrown down, the top of the tent was carried away by the wind, the brother was deafened, and the reindeer were thrown into a panic. On regaining composure, he was not able to round up but very few of the reindeer. All this so worked upon him that he was sick for a long time. In the broken-down forest at one place a hole was formed out of which flowed a little brook into the river Chamba. Through this locality [i.e., where the woods were broken down] formerly went a Tunguse road; now it is abandoned, because it turned out to be all obstructed [by the broken down trees] and impassable, and, besides, for the reason that it caused horror in the Tunguses.
"

Kulik was aware about Evenk's accounts collected by S.V. Obruchev in 1924 (Obruchev, 1925). Obruchev marked that Evenks rejected the fact of the meteorite fall, but were ready to show the area of the forestfall near the Chamba river. Obruchev wrote (translated from Russian with remarks in [...]by A.Ol'khovatov) (Obruchev,1925):

"In the summer of 1924, I was sent by the Geological Committee for research of the r. [river] Podkamennaya Tunguska. During the work, I assumed to attend the place of the meteorite fall. Unfortunately, I failed. (...)
The rumble of the meteorite was heard both in the Teterya and Vanovara factories [trading posts], on the Podkamennaya Tunguska and on the Angara river in all visited by me in 1924 villages from s. [settlement] Dvorets to s. Panovskoe. The rumble was heard in the morning (according to other indications, in lunch ["dinner" - "obed" in Russian] - i.e. about 10 hours). The [window's] glass was trembling in the north side, the objects fell from the shelves, in one case the horse on which they drove, fell. In the Teterya factory [trading post] fiery columns were seen in the north."

The fiery columns seen from Teteya are intriguing, as well as 2 times given.

Also Kulik was informed about Evenk's accounts collected by I.M. Suslov in 1926 before they were published in 1927. In 1927 Kulik reached the area of the forestfall basing on the info.

Detailed investigation of this (Kulikovskii) forestfall started in 1960 and continued for about 2 decades. In the summer of 1961, the expedition of the Committee on Meteorites of the USSR Academy of Sciences, with the participation of the KSE, made a ground survey with trial areas on average on a grid of 2 x 2 km across the entire area of the Kulikovskii forestfall. The size of the trial areas of 0.25-0.50 hectares was selected in such a way that about 100 fallen trees could be taken into account. The location of the trial areas was planned to be as uniform as possible. In cases where there was a choice, preference was given to areas with a better pronounced forestfall (Boyarkina, et al., 1964). This work continued for many years. The data collected was published in a catalog in two parts in 1967, and in 1983. On Fig.1 here is a drawing of the fallen trees directions in the trial areas from (Lyskovskii, 1999). A grid of special coordinates (Fast, et al., 1976) is plotted, a bold dot indicates the intended epicenter of the Tunguska explosion. The points are trial areas, and the vectors from these points are the average directions of the fallen trees on these trial areas. Coordinates (x,y) are given in km. Axis "x" is directed to the magnetic north (see details in (Fast, et al., 1976)). Please pay attention that the trial areas near x=40 km, y~0 km belong to so called Chuvar forestfall (see below).

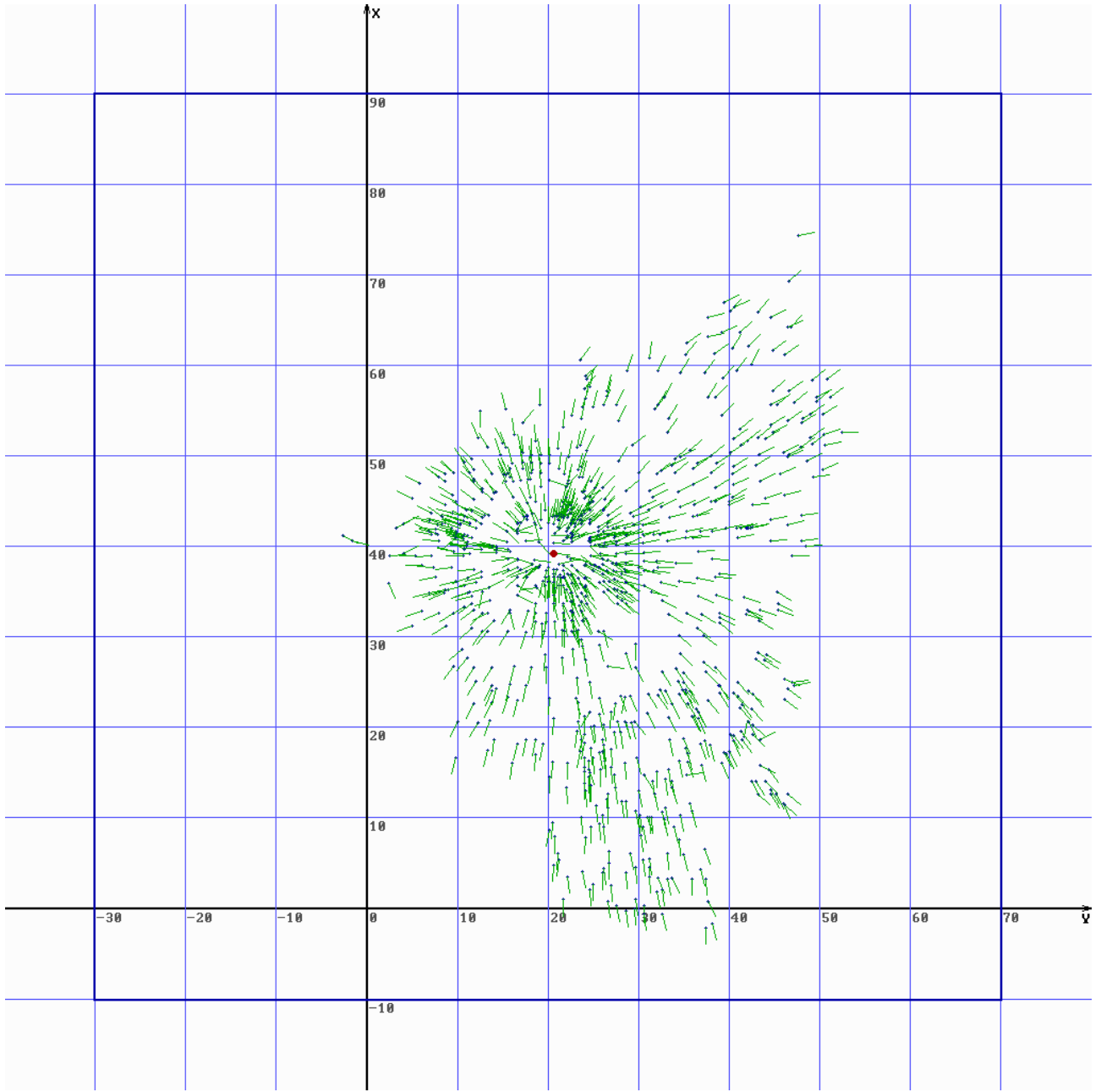

**Fig.1**

Here is a drawing on Fig.2 (Fast, et al.,1976) based on the first part of the catalog in general. The drawing shows smoothed data of the directions of the fallen trees.

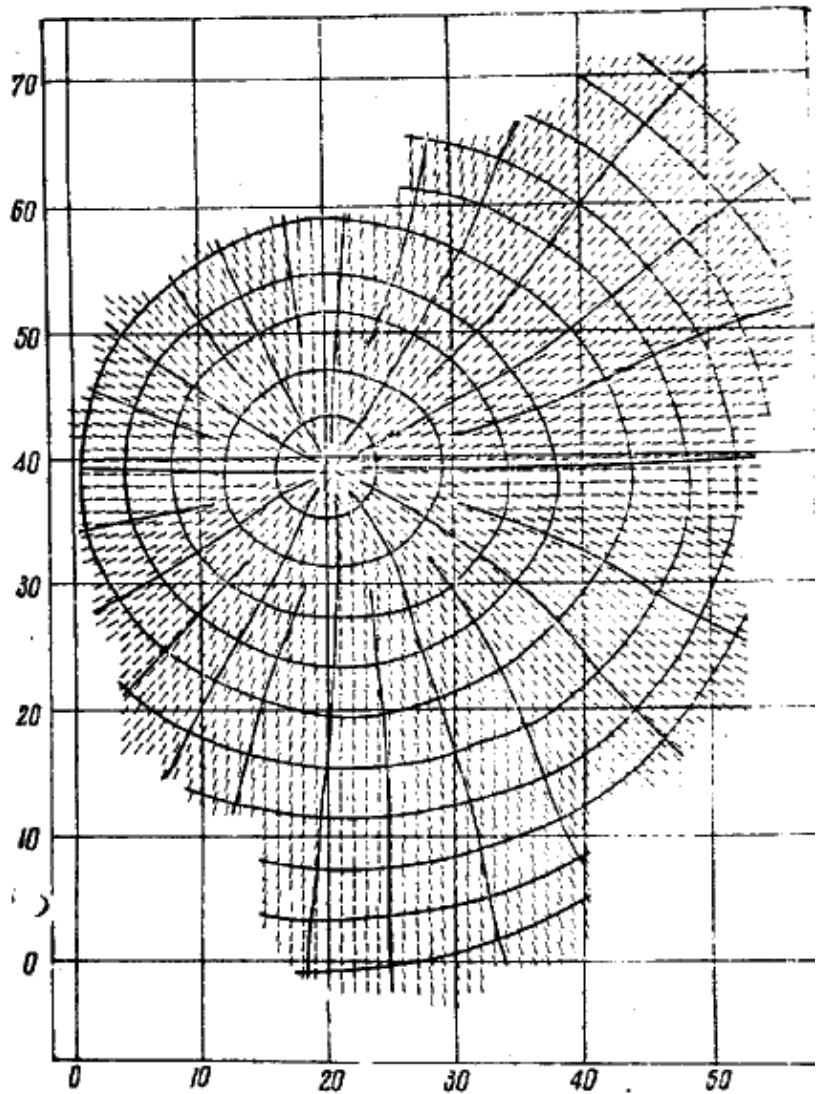

**Fig.2**

Please pay attention that the pattern of the forest-fall on Fig.1, 2 shows only directions of the fallen trees. It does not show a level/degree of the forestfall (i.e. number of trees fallen per square). For example, the forest-fall practically disappears just a few kilometers to the west of the epicenter, and farther to the west just rare fallen trees exist (see also below).

In 2005 Italian researchers with their Russian colleagues presented a new much more complete map. Here is their explanation (Longo, et al., 2005):

"The correspondence between the "kilometre coordinate system" used by Fast and the standard geographical coordinates has never been published. In the new unified catalogue, for each Fast azimuth we give its kilometre and geographical coordinates. The last ones have been obtained by using reference points recognised on the ground. Though part of Fast trial areas data was not used due to the rather poor statistics, the new catalogue includes 1165 azimuths extracted from Fast data [1-2] and published here after the introduction of the necessary corrections. To these data, 80 Anfinogenov azimuths and other 350 obtained from the digitalized photos of the 1938 APS have been added. Thus, the data we used are several times larger than those in Fig. 1 or those considered by Fast to obtain the mentioned TCB trajectory parameters. We have introduced a reliability degree for each trial area averaged azimuth. In Fig. 3, the white, gray and black areas correspond to a high, medium and low reliability, respectively. In the figure, the external frame represents the kilometer coordinates, while the inner - the geographical ones."

In (Longo, 2007) the reliability degree is explained as follows:

"Moreover, we have introduced a reliability degree for each trial area averaged azimuth. The reliability degree has been assigned on the basis of the percentage of singletree azimuths that lay in a sector of 15° centered on the averaged azimuth."

In other words, the reliability degree shows how well-ordered is the forestfall in some place/area. Also it could be speculated that if to propose that Tunguska was an explosion and trees were felled by the air-shock-wave-caused aerial disturbance, then the trees felled by the Tunguska explosion should be fallen in approximately one direction (in some place/area). So if in some area trees fell in approximately one direction then it is a good reason (in the frame of the proposal) to assign them to the Tunguska explosion. Such area would have the high reliability degree. But if the trees lay in various directions in some place, then such an area would have the low reliability degree, as it is not completely clear whether the trees were uprooted by the Tunguska event indeed.

Here is the new map (Fig.3) taken from (Longo, et al., 2005):

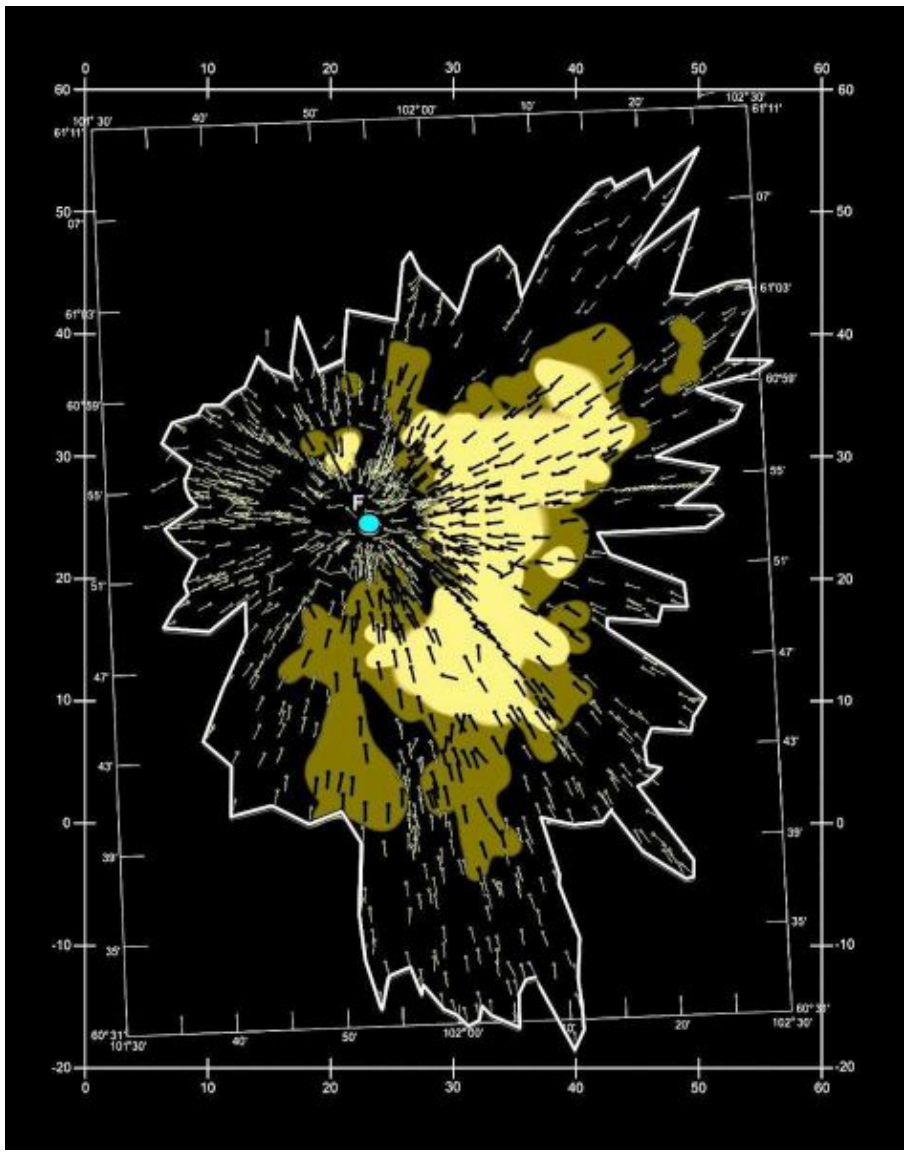

**Fig.3**

A remarkable feature of the map is absence of the high reliability degree to the west of the epicenter. In other words to the west of the epicenter trees were felled by Tunguska rather chaotically or/and their number was small and they were 'dissolved' by trees felled by none-Tunguska reasons. Both variants hint that the Tunguska explosion influence was rather weak to the west (and partly north-west) of the epicenter.

This is in agreement with result obtained by J.F. Anfinogenov and his group. In the mid-1960s J.F. Anfinogenov made a map of completely uprooted area (the map is based on aerial photo-survey conducted no later than 1949). Here it is on Fig.4 (Anfinogenov and Budaeva, 1998):

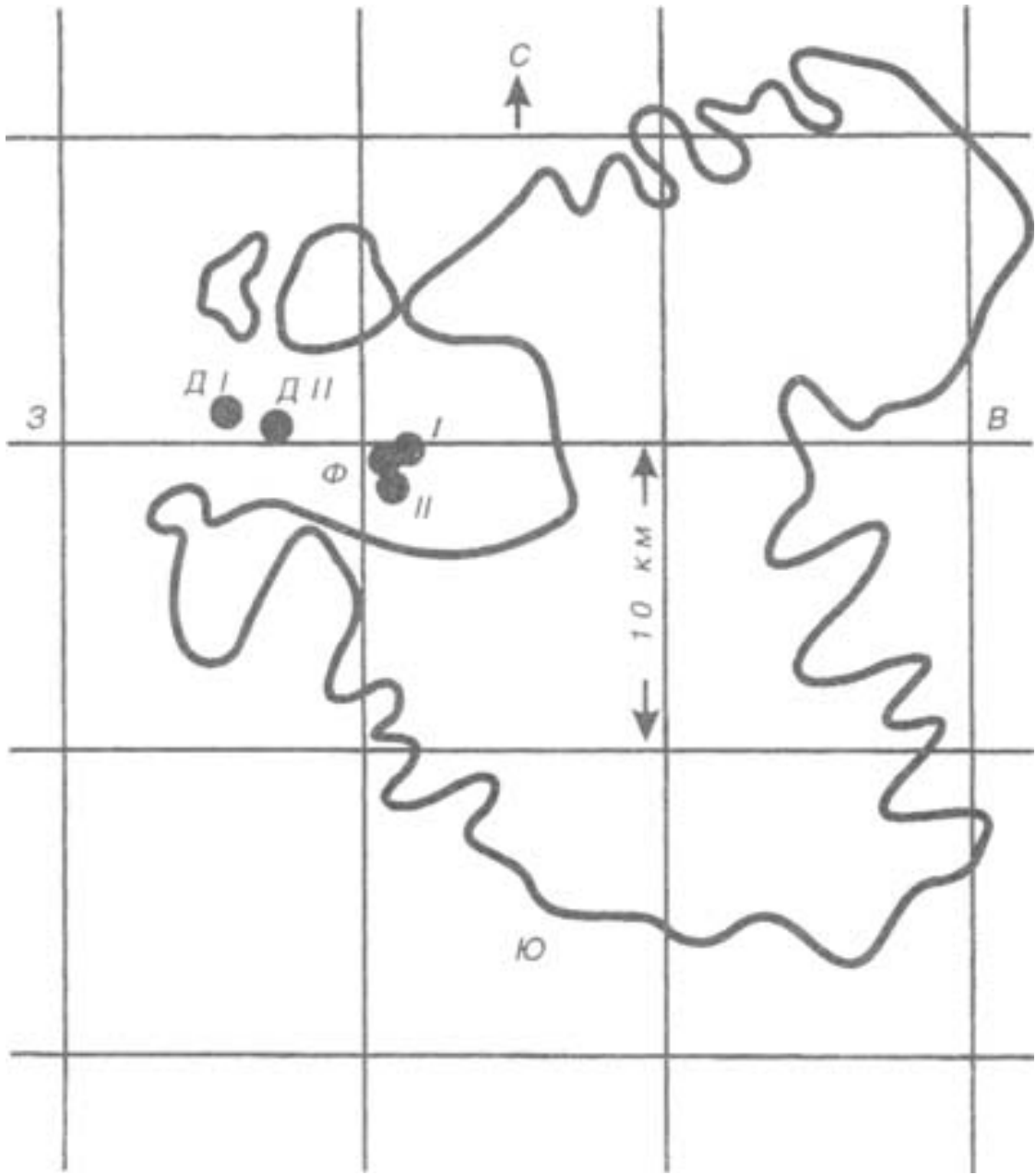

**Fig.4**

The epicenter is marked by a dot near Russian letter looking like a letter 'o'

divided vertically. It is seen from the map that there was no 'total uprooting' to west of the epicenter. It is noteworthy that a couple dozen kilometers to the west of the epicenter there is the Chuvar forestfall in which trees fell in reverse direction (see below).

Please pay attention that as the forestfall catalogs are based in general on sampling in small trial areas separated by a distance of about 2 km. So some "thin" structure (details) of the forestfall can be missed. Regarding small-scale structure - Kulik make a scheme (after getting aero-photos of the central region of the forestfall) with two centers of direction of tree-fall - see Fig.5 –from (Kulik, 1940). Kulik signed the picture in the following way:

> Scheme of the western part of the Great Southern Marsh with two centers of direction of tree-fall and isobaths, plotted according to the work done in 1939. Scale—1 cm = 48 m. The contours of the Great Southern Marsh, of the «islands» with the areas of frozen hilly turf-pits preserved on them, are shown with a thick line. The isobaths of the marsh bottom as taken in 1939 are denoted with a thin line. Straight lines indicate the direction of fallen trees; the centers are denoted with circles.

Since then, this question (reality of these 2 epicenters) rises again from time to time, but there is no final answer to it yet.

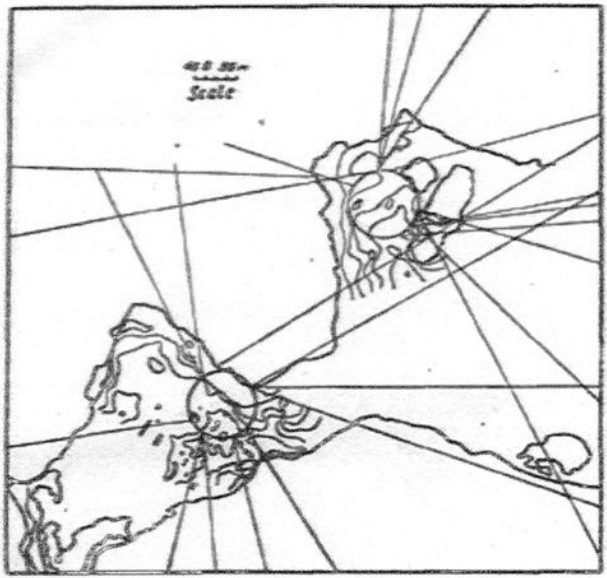

**Fig.5**

It is interesting to note about another article in which a possibility of a second epicenter is stated - here is from an abstract of (Goldine,1998):

"It is found that in addition to main epicentre of the forest destruction region, previously determined, the method indicates another critical point located about 4–6 km to the west of the main epicentre. This feature can be interpreted as a consequence of the flight and destruction of small piece of the Tunguska meteorite."

Another peculiar aspect of the Kulikovskii forestfall was presence of survived trees close to the epicenter. Here is from (Krinov, 1949) (translated from Russian by A. Ol'khovatov):

"On the shores of the Hushmo River, especially to the west of the landing place of the expedition, i.e. upstream, more and more often there are kurtins and groves of growing forests, and already at a distance of just a few kilometers there are significant sections of untouched forests, which are like islands inside continuous forestfall and deadwood. The preservation of these groves is not always clear, since there are no obstacles to the spread of an explosive wave around them. Moreover, sometimes near the growing forest sites on flat areas, there is continuous fallen forest oriented towards the basin, which is located at a distance of 5-8 km north-east. The idea is created that the explosive wave acted far unevenly around the place of the meteorite fall and that not only the terrain only had a protective effect. It was possible to conclude that the explosive wave had a "rays-like" character and, as it were, "snatched" certain sections of the forest, where it produced continuous forestfall or other destruction. Such a "snapping" of individual sites was particularly well observed when viewed by aerial photographs relating to areas, located at a distance of 2-3 km west of the place of the meteorite fall."

Krinov suggested that more developed root system of coastal plants probably contributed to better preservation of the forest on the shores of the Hushmo River. However then he wrote (Krinov, 1949) (translated by Andrei Ol'khovatov with addition in […]):

"From the tops of the northern hills, the author saw the bluish taiga, located immediately behind the Swan Lake [the Cheko lake] and farther to the north of it. As far as it was possible to determine during observations, from a distance of about 6-8 km, this site of the surviving taiga is located on an elevated, no secure place. Therefore, the safety of the forest from the action of an explosive wave in the specified place is completely incomprehensible. <...>

Further, in the indicated areas, the unevenness of the forest collapse - "grabbing" was striking. In some places one could see separate glades, where the forest was tumbled down completely. But right there, nearby, there were areas with a standing growing forest. The contours of the sites with the fall out of the forest are formless and there is no way to find them in relation to the Southern Swamp."

Similar questions were raised by various researchers. For example, here is what was written in 1964 regarding individual trees (Boyarkina, et al., 1964) (translated by A. Ol'khovatov):

"Individual trees are distributed everywhere in the northern, western and eastern directions.

Despite careful attempts, no old trees were found to be confined to the relief. They are also on the inner and outer slopes. It is impossible to talk about shielding them. The preservation of old trees is probably due to biological reasons: strong roots, good, moist soil, the long absence of fire, i.e. the strength of the forest.

In the northern direction, the old forest generally comes close to the epicenter."

And here is about the whole groves of trees - from (Kharuk, et al., 2006) (translated by Andrei Ol'khovatov):

"As source data in the work aerial photography materials were used ... made in July 1938, as well as aerial photography materials on the same territory conducted on July 26, 1999 .... <...>

Two areas were selected in order to reduce temporary and resource costs processing of the images. The choice of these areas was due to that only they had overlap of large-scale (1:10 000) photos of 1938 and 1999. Analysis of these areas in the photographs of 1938 revealed the following anomaly: in the area-2 almost completely absent forestfall, also not noticed "telegraph forest" ( trunks of trees, completely devoid of branches, but not fallen by the shock wave). There are also no visible marks of foresfire.

A somewhat different picture is observed in the area-1, where there are fallen trees, and tree trunks without crown. <...>

As noted earlier, the analysis of the studied areas in 1938-photos shows anomaly in the forestfall. Considering the scale of phenomenon as a whole, and small, about 600 meters, the distance between the areas, this anomaly cannot be explained by weakening shock wave with retreating from the center of the explosion, if to take the epicenter of the explosion center, calculated by the Fast's catalog of the forestfall. <...>

When analyzing the relief, on the topographic map of 1: 100000 is seen that the area-1 is located in the southeast, gentle (3-4 °) slope, while the area-2 on the north, more steep (7-8 °) slope. The northern slope can partly explain the lack of fires in the area-2 and associated damage of the forest, because higher humidity level at this time of the year. It could be an obstacle for the spread of the lower fire, as well as for inflammation of the litter from the thermal radiation of the explosion, which undoubtedly had place in the catastrophe area.

As for the direction of the shock wave, then considering that the majority of researchers give an estimate of the height of the explosion in 5-7 km, then minor, not more than 50m, elements the relief could not play a significant role in the case of one, central or volumetric, air explosion. But if to allow a lower explosion option, then the relief could be partial or fully supress the shock wave."

The considered areas were close (witnin several km) from the Fast's epicenter. Also please pay attention that the statement "...inflammation of the litter from the thermal radiation of the explosion, which undoubtedly had place in the catastrophe area" (while sounding plausible) is not proved.

It is important to note that there were many survived groves of trees near the epicenter. Here is some details (Boyarkina, et. al., 1964) (translated by A. Ol'khovatov):

"Of considerable interest is the presence and distribution of living old trees. The epicentral zone contains a significant number of individual old trees and entire groves, which was discovered back in 1959-1960 [4, 12].

In the diagram [8, FIG. 6] all notable groves of old trees are marked. The largest of them are: on the slope of the Wulfing (500 X 700 m), on the southern shore of the Southern Swamp (200 X 700 m), an array on the Western peat bog, a 200 X 300 m, groves to the east of the swamp, etc."

The grove on the slope of the Wulfing mountain (or a hill – its height is below 500 m) is probably the largest one. Here is how it is described in (Romeiko, 2006) (translated by A. Ol'khovatov):

"Traveling along the edge of the swamp, we will come across small groves of old trees more than once. Researchers have repeatedly drawn attention to this feature of the epicenter of the explosion. It seems that the explosion fragmentally felled the taiga, leaving some areas completely intact. So, for example, it was with a grove of old trees standing under the Wulfing Mountain."

Here is what I.K. Doroshin wrote about the Wulfing grove (Doroshin, 2005) (translated by A. Ol'khovatov):

"...at least one grove consisting of larch, spruce and cedar (near the Wulfing mountain). In such places, the trees either did not lose their crowns at all, or the loss was minimal. There are either no traces of the 1908 fire here at all (a grove near the Wulfing mountain), or there are traces of a grass-roots fire of varying intensity;... "

The Wulfing grove is situated within ~3-4 km to the north-west from the epicenter.
Any interpretation of the 1908 Tunguska event should explain these peculiarities.

And here is one more important factor pointed out by Prof. Evgenii Ivanovich Vasil'ev (the Volgograd University). In 2008 the graduation degree of the student Olga Bolshakova (the Volgograd University) was devoted to checking the results of the calculations of Aleksei Vasil'evich Zolotov (Zolotov, 1969). According to research by Zolotov (Zolotov, 1969), concentrations of the fallen trees direction's intersections in the northeast part of the Kulikovskii forestfall are inside this (i.e. northeast) part (see Fig.6). On Fig.6 red lines mean directions of the fallen trees.

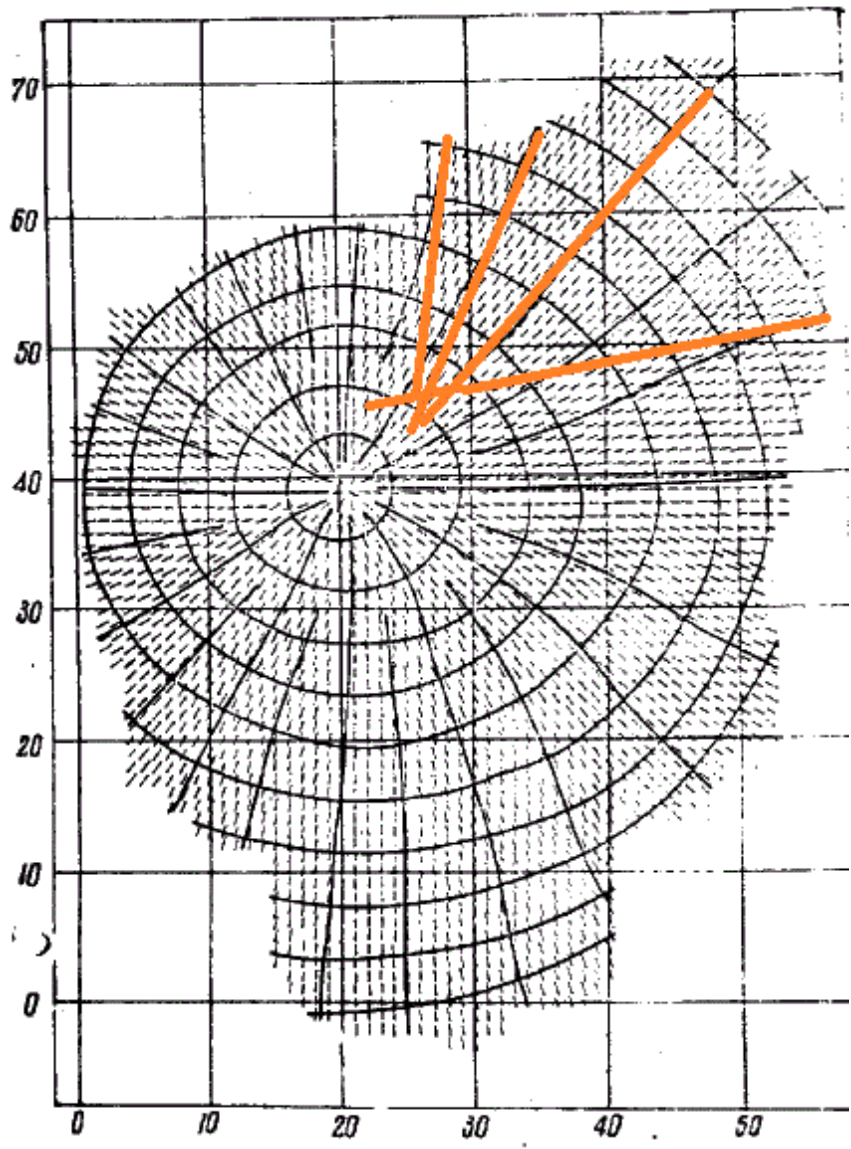

**Fig.6**

Also concentrations of the fallen trees direction's intersections in the southeast part of the forestfall are inside the southeast part
 (see http://tunguska.tsc.ru/ru/science/1/zol/4/12/ for details of the Zolotov's calculations).

To be more precise, it should not be about the east-west line, but about the axis of symmetry of the Kulikovskii forestfall, which, however, are very close. Therefore, to explain the essence of Zolotov's calculations, such a simplification is permissible.

According to Bolshakova, this effect is pronounced for trees most distant from the epicenter. Thus, the directions of the fallen trees in the outer parts of the Kulikovskii forestfall "wings" deviate from the radial (from the epicenter), and have a divergent/expanding character.

Zolotov made his calculations with initial limited data of the Kulikovskii forestfall, which contained data of 338 trial-areas. Bolshakova checked and confirmed the Zolotov's result using much complete dataset containing data of 950 trial-areas.

Bolshakova applied the results obtained to the analysis of the Tunguska event calculations presented in the news release:
https://newsreleases.sandia.gov/releases/2007/asteroid.html
Here is the Bolshakova's outcome (translated by A. Ol'khovatov):

> "According to the results of this article[6], it turns out that the stronger section of the wave has a flat shape and a much larger radius curvature. Therefore, the center of the trees felled by this section of the front should be on the other side of the axis of symmetry. The performed analysis shows that the opposite picture is observed. Therefore there are doubts about the reliability of the model in which the fall of the forest is carried out by the blast wave. And it is quite possible that the mechanism of falling out of the forest in the wings were somehow different."

Let us explain the situation with more details. In the news release:
https://newsreleases.sandia.gov/releases/2007/asteroid.html
there is a movie-clip "Movie 6": (http://www.sandia.gov/videos2007/2007-6514PFvmag-tun2.4.mpg). Here is explanation in the news release for the movie 6:

> "Map view of blast zone from 3-D simulation of a 5 megaton explosion. Axes are labeled in centimeters, and colors indicate wind speed. Expanding oblong shape is the blast wave moving along the surface, blowing down trees with wind speeds decreasing from high hurricane force of 60 m/s (magenta) to below 20 m/s (yellow). Because the fireball stops at high altitude, there is no blast furnace zone near the epicenter and trees remain standing as observed at Tunugska."

Let's look at a snapshot of the "Movie 6" in which directions of the falling trees are added with red lines (Fig.7).

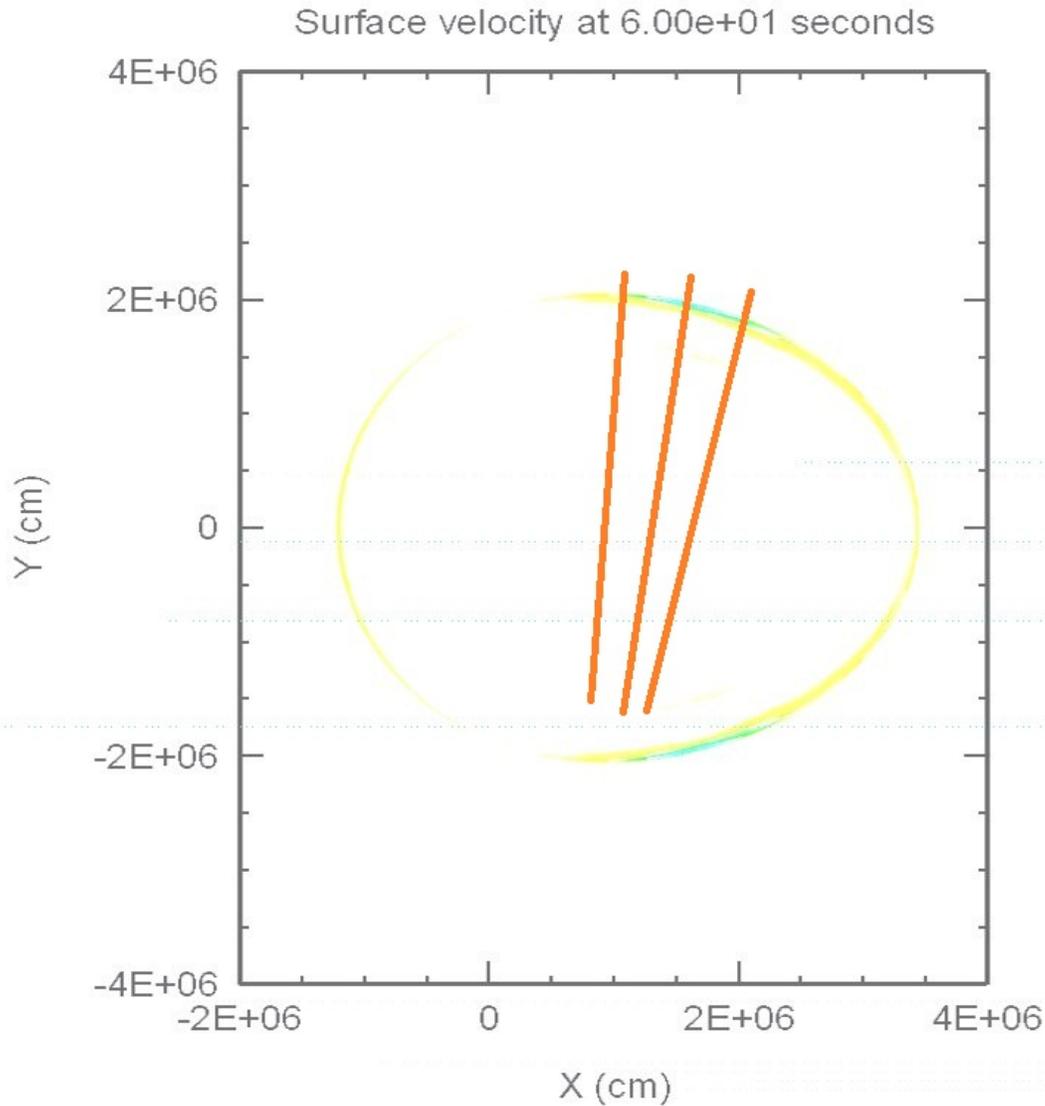

**Fig.7**

The snapshot corresponds to a moment when (according to the interpretation of the Tunguska event presented in the news release) the outer regions of the Kulikovskii forestfall are created.

It is seen on Fig.7 that as Bolshakova wrote:

"…it turns out that the stronger section of the wave has a flat shape and a much larger radius curvature. Therefore, the center of the trees felled by this section of the front should be on the other side of the axis of symmetry."

So the interpretation of the Tunguska event presented in the news release does not conform to the structure of the Kulikovskii forestfall.

## 3. Some other forestfalls

**The damage at Ognia.** Historically the first place of the alleged (by L.A. Kulik) "meteorite fall" was near the river Ognia. Engineer V.P. Gundobin wrote to Kulik in 1924 about the Evenk's story that something flew from the sky, put down trees, then a forestfire started. Also Gundobin wrote that he met the Evenk's prince Dushinchi, who confirmed that at "that time" there was a forestfire, and a mountain's collapse at the river Ognia.

This place is about ~55-60 km (azimuth about 128 degrees ) from the epicenter. A possibility of the forestfall's small patches seems reasonable. Indeed according to Kulik (Kulik, 1937):

> "The investigation brought to light a continuous, eccentric, radial "windfall" in a huge area of radius about 30 km., which extended on some of the hills to a distance of 60 km. and even farther. Observers relate that individual trees were thrown down on the hills in the neighborhood even of Vanovara."

Krinov confirmed (Krinov, 1949) that the first traces of the taiga's damage were near Vanovara (translated by A. Ol'khovatov, factory here means "trading post" ):

> "If you get acquainted with the attached map, you can see further that the first traces of the action of the explosive wave on the periphery of the region are noticed almost at the Vanovara factory itself, where ( especially along the riverbed of the Chambe River, starting from its mouth )  trees with broken tops growing on the shores are observed. Such trees sometimes are met alone, sometimes in groups with several trunks located nearby."

So it is possible to admit presence of some minor forestfall near Ognia.
Anyway, the statement about a forestfire at such distance from the epicenter is remarkable.

**The Ayan forestfall.** Vasily Mikhailovich Arbatsky from the settlement of Krasnoyarovo (~57.3 N, ~107.5 E) wrote in 1922 about the event (Vasil'ev et al., 1981) (translated by A. Ol'khovatov):

> "The phenomenon was on June 17, 1908, starting at 10 a.m. It was observed partly from the field and from the place of residence of the village of Krasnoyarovo. The weather that day was clear, but the air for breathing was heavy. From the very beginning, what was heard or what attracted the attention of this phenomenon were sounds similar to the sounds of guns with interruptions in the direction from west to north. At first, the sounds were heard more often, after 5 minutes, and then less and less often. The blows lasted for 1.5 hours. It was impossible to count the number of blows. In this direction, where the blows were heard, the sky was slightly darker. The impact of this phenomenon on both birds and animals did not produce anything special, People had various opinions. Some expected a rain cloud, the blows were recognized as thunderous. In the same year and a month, the day is unknown, from the river Tunguska in 15 versts on the river Ayan turned out to be a big strip of the uprooted forest (together with roots) on the road for 1 versts, and to the sides - unknown. There are many eyewitnesses of this, residents of the village of Krasnoyarovo, who are engaged/working in Tunguska, among whom I was."

In Siberia, several rivers are named Ayan, and not far from the village of Krasnoyarovo, two nearby rivers flow into the Nizhnyaya Tunguska: Ayan and Ayan Pervyi ("the first"). However, in any case, the choice of "just Ayan" or "Ayan 1" does not have a fundamental effect on the distance from the epicenter. It is approximately ~450 km and an azimuth of about 140 degrees.

It is also interesting to note in the account a slight darkening of the sky in the direction of thunderclaps and their long duration. For comparison, an experienced and respected observer of the meteorological station in Kirensk (60 km to the NE from Krasnoyarovo), G. K. Kulesh, wrote in 1908, 6 days after the event that the phenomenon lasted from about 7.15 am to 8 am (Vasil'ev, et al., 1981). It is noteworthy that in this case, too, there is a similar discrepancy in the time of the beginning of event, noted by Obruchev (see earlier).

Although the exact date of (appearence) the Ayan forestfall is unknown, there is a hint that June 30 (the Gregorian calendar) is the most likely. Indeed according to the Kirensk weather station, located about 60-70 km east of the Ayan forestfall site, a strong wind and a thunderstorm were observed in the second half of June 30. This is the only case of registering a strong wind at the Kirensk weather station for the entire summer of 1908 - see also (Ol'khovatov, 2020b). By the way the Kirensk meteostation is known for its high data reliability, thanks to its observer G. K. Kulesh.

According to the Kirensk meteostation log the strong wind and the thunderstorm occured between 1 pm and 9 pm (times are local) on June 30, 1908. At 7 am the meteostation registered "no wind", at 1 pm a wind from N with velocity 2 m/s, at 9 pm - a wind from SE, 8 m/s.

Also there is a note in the "Sibir'" newspaper of July 2 (the Julian calendar) by S.K. (probably S. Kulesh) about events of June 30 which stated that (translated by A. Ol'khovatov):

> "About 2 pm between Kirensk and Korelino (closer to Kirensk) on the same day there was an ordinary thunderstorm with heavy rain and hail."

The mentioned village Korelino (Karelino) was about 25 km from Kirensk.

By the way, by chance or not, the direction of sound movement heard in Krasnoyarovo (from west to north) roughly corresponds to the movement of the sound source from the Ayan River towards Kirensk.

G.K. Kulesh was at home during the event, so he collected and reported what others had seen (Vasil'ev, et al., 1981) (translated by A. Ol'khovatov):

> "On June 17 (old calendar), a phenomenon was observed on the NW from Kirensk, which lasted from about 7 o'clock. 15 min. to 8 h. in the morning. I did not get to observe it, since I, after recording the meteorological instruments, sat down to work. I heard muffled sounds, but I took them for volleys of gunfire on the military field beyond the Kirenga River. When I finished my work, I looked at the barograph tape and to my surprise noticed a line next to the line made at 7 o'clock. in the morning. This surprised me, because during the continuation of work I did not get up from my seat, the whole family was asleep, and no one entered the room."

G.K. Kulesh described the event in the following way (Vasil'ev, et al., 1981)

(translated by A. Ol'khovatov, sazhen = 2m 13cm):

> "Here's what happened (I'm passing on the essence of the eyewitness stories). At 7.15 am a pillar of fire appeared on NW, four sazhens in diameter, in the form of a spear. When the pillar disappeared, five strong short thunders were heard, as if from a cannon, quickly and clearly following one another; then a thick cloud appeared in this place. After 15 minutes, the same blows were heard again, after another 15 minutes, it was also repeated. <...>
> There were also peasants from the village of Korelino, which lies 20 versts from Kirensk on the nearest Tunguska, they reported that they had a strong shaking of the soil, so that windows were broken in the houses."

It is interesting to note that the Korelino village lay to the northwest of Kirensk. As G.K. Kulesh did not see the "pillar of fire" himself, so he had to convey impressions of the eyewitnesses regarding the diameter of the "spear".

The "spear" is not the only luminous phenomenon reported by residents of the region around Kirensk. However, the discussion of the observed light phenomena is beyond the scope of this paper.

In 1997 a new account from Kirensk by Ivan Suvorov (who was in open air) was presented by V.A. Bronshten (Bronshten, 1997). Here is a fragment of it (translated by A. Ol'khovatov):

> "Ivan liked to get up early and do jogging in one verst. The morning of June 30, 1908 was no exception. This morning was cloudless, the sun was shining brightly with no wind at all. Suddenly, Ivan's attention was attracted by an ever-increasing noise, which, as it seemed to him, came from the south-eastern side of the sky. Nothing like this was felt from the east, north, or west. The sound was getting closer. "All this began," wrote Ivan Suvorov, "according to my watch, verified the day before at the Kirensk post office, at 6:58 local time. Gradually, the approaching sound source began to be listened to from the south-south-west side and moved to the west-north-west direction, which coincided with the fiery column that shot up into the sky at 7:15 in the morning.""

Then V.A. Bronshten commented the Ivan's account (translated by A. Ol'khovatov):

"What surprises us about these statements? First of all, the time of the beginning of the audibility of the abnormal sound is 6 hours 58 minutes, while the pillar of fire shot up, in full agreement with other definitions, at 7 hours 15 minutes. The Tunguska bolide could not fly, making a sound, for 17 minutes. During this time, at a speed of 30 km/sec, it would have flown 30,000 km, that is, at 6 hours and 58 minutes, it was far beyond the atmosphere and could not make any sounds. This means that this moment does not refer to the beginning of the appearance of sound, but to some other event, for example, to Ivan's exit from the house.

The correct indication of the moment of the explosion makes us reject all other possible assumptions: for example, that Ivan's clock was 17 minutes behind in a day, or that the local time of Kirensk was very different from the local times of other points. Moreover, in the same Kirensk, the director of the meteorological station, G. K. Kulesh, recorded the arrival of an air wave ( i.e. the same sounds ) after 7 o'clock according to the barograph readings.

Ivan also inaccurately recorded the direction from which the sounds came. The Tunguska bolide flew, according to the most accurate definitions, to the north of Kirensk. The closest point of the trajectory was from it to the northeast. Then the bolide moved to the north and finally, to the northwest.

As E. L. Krinov reports in his book "The Tunguska Meteorite" (Moscow: USSR Academy of Sciences, 1949, p. 54) that many eyewitnesses later claimed that they heard the sound before they saw the bolide ( which in fact could not be ). Apparently, this is some kind of peculiarity of inexperienced observers who reported what they saw much later, several years after the event."

The words of V.A. Bronshten speak for themselves, and the only thing the author would like to add to the words of Bronshten, is the fact that if Ivan Suvorov could hear the air-wave which was detected by the barograph (amplitude of the latter, according to I. S. Astapovich ([Astapowitsch](Astapowitsch), 1940) was 1.1 mm Hg., and it came only in 7 h 48 min of the morning), then he would (and all the witnesses from Kirensk) report about very different things. The fact is that the air wave of sound frequencies with amplitude of 1.1 mm Hg. approximately corresponds to the pain threshold. The barograph recorded exceptionally slow, inaudible pressure changes, practically without reacting to sound.

**The Sulomay forestfall**. In 2008 Tunguska eyewitness accounts collected by the Russian ethnographer Sev'yan I. Vainshtein were first published in Russian (Ol'khovatov, 2020a). These had been collected during his 1948 expedition to the settlement of Sulomay (61.6°N, 91.2°E) which is situated about 577 km west of the Tunguska epicenter. On the morning of June 30, 1908 the bright fireball was seen, accompanied by thunderous sounds. The earth began to tremble, and a strong wind swept in. The wind uprooted tall trees in the taiga and collapsed tents; women and children cried and shouted (Ol'khovatov, 2020a). Unfortunately details of the forestfall remain unknown. Anyway the accounts collected by Vainshtein were included in the second (electronic) edition of "Katalog Pokazanii Ochevidtsev Tungusskogo Padeniya" compiled by L.E. Epiktetova in 2018 ( http://tunguska.tsc.ru/ru/science/1/eyewitness/ ).

**The Ket's forestfalls.** In 1948 there was an article by P. L. Dravert "Burelom i ozshog lesa v basseine reki Keti." In this article (on the basis of a visit by a geologist to the basin of the river Ket in 1932) it was suggested the fall of two fragments of the Tunguska meteorite in the basin of the Ket river (Vasil'ev, et al.,1963).
      In the late 1950s G.F. Plekhanov said (Erokhovets, 1960) (translated by A. Ol'khovatov):

> "In the upper reaches of the river Ket, north of Tomsk, there is an area of the forestfall. Moreover, as they say, this forestfall is similar to the destruction in the Tunguska catastrophe area. Some windfalls relating to about 1908 are on the river Korda. They are described by Dravert ... So, I took the sake of curiosity took a globe and connected these points from the river Ket to the place of falling the Tunguska meteorite. It turned out a rather straight line. What is it - randomicity? May be. But the randomicity is strange and interesting. Need to check? Necessarily."

      The Ket river basin was investigated in 1960 by an expedition (Vasil'ev, et al.,1963). Several forestfalls were discovered (dimensions can be as: length ~40 km and more, and ~4 km in width). The forestfalls were of strip-like character. Vasil'ev with colleagues concludes (Vasil'ev, et al., 1963) (translated by A. Ol'khovatov):

> "As for the causes of the windfall in the Ket River basin, then, judging by old-timers, it is associated with two hurricanes, the first of which was between 1906 and 1912, and the second - between 1921 and 1930. <...> Much later on the site of the windfall there was a forestfire, which is connected with those traces of burn on trees that are visible on the fallen trunks even nowadays. Thus, they are not related directly to the event, which resulted in the windfall.

Considering the fact that the direction of the tree-falls coincides with the prevailing direction of the winds in this area, it can be assumed that the specified windfall has nothing to do with the flight or fall of the Tunguska meteorite. In any case, the strip-like fall of trees is typical for windfalls, the causes of which are strong windstorms."

Interestingly, that at least one forestfall occurred between 1906 and 1912 (which includes 1908...), and also that direction of the forestfall is "looking" to the side of the Kulikovskii forestfall, which is situated about 900 km away to the east. Randomicity? May be...

**The Eastern forestfalls.** G.P. Kolobkova wrote from Vanavara to G.F. Plekhanov in 1959 ( http://tunguska.tsc.ru/ru/science/mat/oche/31-60/d-039/ ) that local hunters talk about powerful forestfalls on Jelindukon.
One of the hunters was a prominent local resident - Andrei Ivanovich Jenkoul. In 1974 N.V. Vasil'ev talked in Vanavara with him (a local journalist took part in the talk too). Andrei Jenkoul stated that on ridges in the areas of rivers Paiga, Jelindukon and Segochamba there are forestfalls, caused by the meteorite. Also he considered as a "meteoritic one" a forestfall on Chuvar (see later in this paper), and stated that however, there are no meteorite forestfalls farther to Mutoray.

A few words about Andrei Ivanovich Jenkoul. Vasil'ev wrote that A.I. Jenkoul is about 60 years old (officially born in 1917), quite educated, sociable, during the war - a lieutenant, commander of a group of snipers. In 1958 he was a guide of the expedition to Tunguska epicenter by Committee on Meteorites of the USSR Academy of Sciences. A.I. Jenkoul was a hunter at the time of the talk with N.V. Vasil'ev. A.I. Jenkoul is mentioned positively in a number of publications on Tunguska. A.I. Jenkoul left this world in 1992...

On web-page ( http://tunguska.tsc.ru/ru/science/mat/oche/92-120/092/ ) there is some additional information. In 2001, Doroshin I. K. and Krivyakov S. V. went from the mouth of the Diergun river up its valley about 15 km. At the top of one mountain, they found a directional forestfall. However, in their opinion, the forestfall was caused by strong winds, because location and shape of the ridge contributes to the strengthening of the wind force at the top with easterly winds. All the trees standing close to the top of the dike have a slope to the west, both young and old. With the natural death of trees, they will fall from the top to the west, which is the case. Both the old and the young forestfall here lies with the peaks to the west. As soon as the mountain's slope became more gentle, the directional forestfall disappeared. The age of fallen trees is very different, from ancient to modern (the assessment is made according to the degree of weathering of the wood). A photo of the forestfall can be seen here:
http://tunguska.tsc.ru/ru/cae/photo/2000/2001/nature/9453/

Approximate position of the Diergun forestfall is 60.6 N, 103.6 E. In other words, the discovered forestfall is about 100 km away (at azimuth 109 degrees) from the epicenter. However the Evenk hunters talked about other rivers than the Diergun river (the Diergun River is a tributary of the Jelindukon River).

The above-mentioned small expedition is one of the few expeditions (related to Tunguska research) that visited this region. Due to the inaccessibility of this region, it has been little explored.

In the summer of 1960, by means of aerial visual observations from a helicopter, members of the KSE examined the basins of the rivers: Tatere, Jelindukon, Segochamba, Bolshaya Yerema, Khuga, and the upper reaches of the Yuzhnaya Chunya. A number of forestfalls were found that were classified as wind-driven. Here is what V. K. Zhuravlev and his co-authors wrote in (Zhuravlev, et al., 1963) (translated by A. Ol'khovatov):

"In 1960, a number of overflights of the proposed area of the Eastern forestfall were undertaken by an airplane and a helicopter. The plan of the survey of the area included the search for a forestfall in the area of the watersheds of the basins of Jelindukon and Kulinda, Segochamba and Bolshaya Yerema, Yeremakan and Altyb. In addition, a helicopter flight was made on the route Vanavara-the upper reaches of the Bolshaya Yerema-Jelindukon (in the middle stream) - Bolshaya Yerema (15 km downstream) - 60°N. — Vanavara. All routes were made taking into account the range of the MI-4 helicopter and the YAK-12 airplane.
Observations in flight on route No. 1 were conducted from the airplane by G. F. Plekhanov and V. K. Zhuravlev. In the upper reaches of the Yuzhnaya Chunya, no signs of a forestfall were found within the route. In a number of places along the course of the Jelindukon River, as well as in the Segochamba River valley, there are areas of chaotic forest fallout that are not connected to each other. As a rule, these sites fall on sites of recent forestfires. Their appearance differs from the picture of the Kulikovskii forestfall, which is characterized by a clearly expressed orientation. Observations from a helicopter in flight along route No. 2 were conducted by V. A. Koshelev and G. F. Plekhanov. In the upper reaches of the Hugi River and between Kulinda and Segochamba, a disorderly forestfall was discovered, apparently located on the old ashes.

Observations in route No. 3 were carried out by V. I. Kolesnikov, V. V. Milchevsky, G. G. Ter-Minos'yan from an airplane. An aerial visual examination of the burnt-out place located approximately at 105°30'E, and 60°15'N, showed that it was caused by an ordinary taiga fire (the presence of a large number of dry burnt trees, the absence of directional forestfall), the examination of the burnt-out place located in the area of 104°10'E, and 60°15'N leads to the same conclusion.

In our opinion, a large mass of young forest of about 50 years of age found to the north of Lake B. Eremakanskoe (at the source of the Eremakan River), in the interfluve of the Pravyi Altyb and Eremakan rivers, may be of particular interest. The observation conditions did not make it possible to establish the presence of a mass forestfall in this area, but there are fallen trees in the area of this young forest. The contour of this area is indented, but in general it is elongated to the southeast."

Approximate position of the latter "suspected" forestfall is about 60.8°N, and 104.7°E, i.e. at distance about 150 km ( at azimuth ~93° ) from the epicenter.

But in 1960 due to the inaccessibility of the places of possible "meteorite" forestfalls, KSE focused on the research of the Kulikovskii forestfall. At the end of the 1960s, when many years of attempts to find the substance of the alleged "Tunguska spacebody" were unsuccessful, some interest in the research of other possible "meteorite" forestfalls was resumed. So the research of the Chuvar ("Western") forestfall was resumed (see below), and in the mid-1970s small expeditions were organized in search of possible Eastern forestfalls. Alyona Petrovna Boyarkina led the "eastern" expeditions. Unfortunately, there are practically no materials about the results of these expeditions left. A rare exception is an article by L. Kabanova in the newspsper "Molodoi Leninets" (Tomsk) of Oct.15, 1974 ( http://tunguska.tsc.ru/ru/lyrics/periodics/70/1974/mollen1/ )
where Boyarkina stated the following (translated by A. Ol'khovatov):

"On July 18, in the upper reaches of the Paiga River and the Yuzhnaya Chunya, we found a forestfall in large areas of the taiga, very similar in appearance to Kulikovskii. We saw the upturned forest, with roots, like frozen hands. All the time we were wondering what the nature of this forestfall was. Wind-driven? After all, the direction of the fall really coincides with the wind rose. Or is it an echo of the shock wave from the explosion of the Tunguska meteorite, superimposed on the wind and determined such a mighty spread of taiga giants? All this will be clarified by further research."

The route of the 1974 "eastern" expedition was about ~80-100 km to the east, NE, and N from the epicenter.

A few words about results can be found in (Vasil'ev, et al., 1981)(translated by A. Ol'khovatov):

As for the aerial visual observations in the Tatere River basin, they, first of all, confirmed the numerous testimonies of old-timers that in the interfluve of Tatere and Yuzhnaya Chunya in the late 20s and later, it was repeatedly passed by powerful (including rare in this area upper) fires and it is a chain of mutually overlapping burn-out areas of different ages. Against this background, it is currently an unrealistic task to identify the forestfall of 1908 in an aerovisual way. In 1975, an expeditionary detachment under the leadership of A. P. Boyarkina made a walking route along the Yuktinskaya road up to the Yuzhnaya Chunya in order to survey the proposed route of the V. Ya. Shishkov's caravan. No large regions of fallen forest that could be confidently attributed to the beginning of our century were found, although in several places the detachment of A. P. Boyarkina came across old, about sixty years ago, strip forestfalls, oriented by the wind rose to NE. Most likely, these are traces of ordinary wind-driven forestfalls that are not related to the issue under study.
Despite all the above, it would be wrong to consider the question of the "Eastern forestfall" removed from the agenda. In combination with the version about the "Pits" on the Yuzhnaya Chunya, it seems quite likely that as a result of the fall of the Tunguska meteorite, some destruction really took place on the interfluve of the Tatere-Yuzhnaya Chunya. However, the verification of this assumption can be made only by labor-intensive ground route work and dendrochronological studies focused on dating of possible forestfalls."

As it was said, unfortunately, there are practically no materials of the results of the expeditions under the leadership of Boyarkina regarding alleged Eastern forestfall[s]. A rare exception is the scheme and a short explanation in this article (Pasechnik, 1986):

"During the expedition of the Tomsk University, the areas of epicenters N2

and 3 were examined and their location was clarified ( oral report by A. P. Boyarkina)."

The scheme is shown on Fig.8 from (Pasechnik, 1986), where it is provided with the following signature:

"The scheme of the known (N1) and newly established (N2,3) possible epicenters of the Tunguska meteorite explosion. The arrow shows the direction of the meteorite's flight path according to [21]."

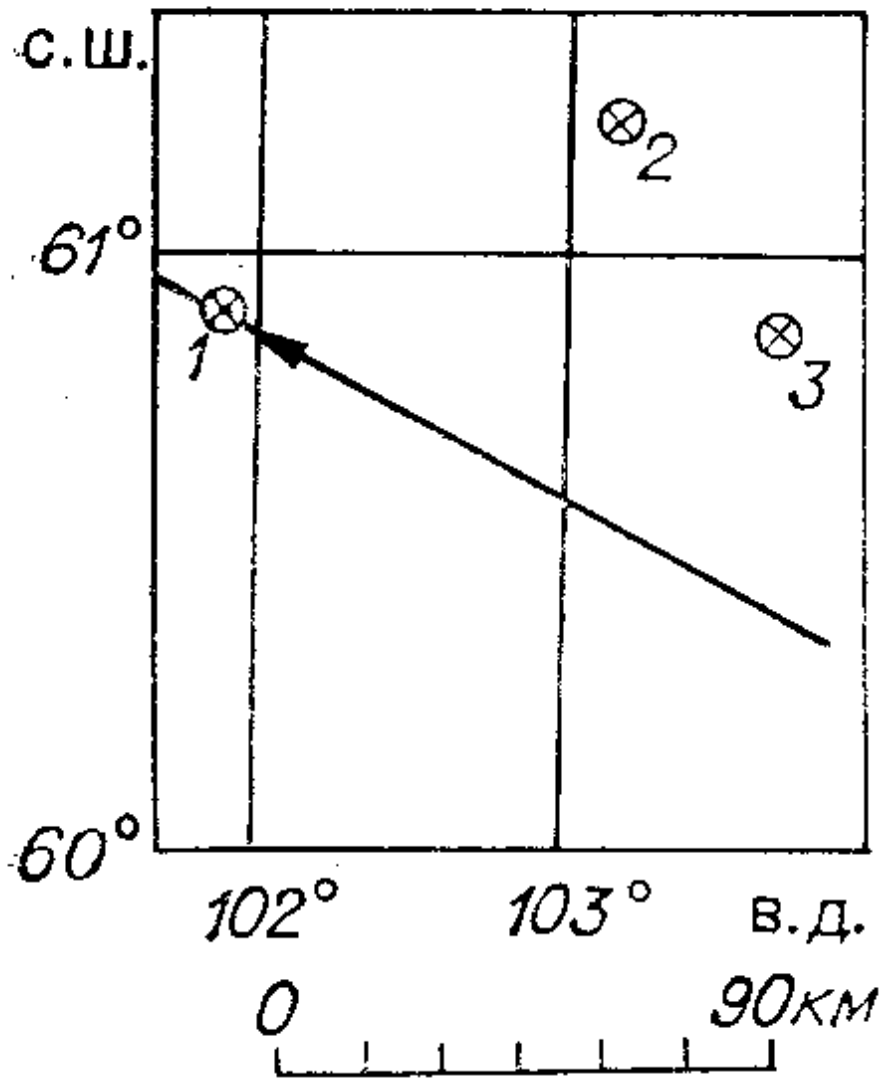

**Fig.8**

Unfortunately since that time no new data appear.

**The Taseeva forestfall**. Here is from (Vasil'ev, et al., 1981) (translated by A. Ol'khovatov):

"A letter of geologist S. A. Khersonsky dated 26/XII-1965 to the Committee on Meteorites of the USSR Academy of Sciences.
S. A. Khersonsky in 1932, as a student-intern, participated in the work of the Angara forest management expedition as part of the cartographic and geographical party. Several detachments were moving through the taiga, making eye-measuring surveys, describing outcrops, collecting rock samples, describing vegetation, etc.
S. A. Khersonsky with two workers and an expeditionary horse was moving along the route along the watershed between the basins of the Chuna and Biryusa rivers. "I had to make my way for six days... through forests and burnt-out areas, among fallen trunks in a chaotic mess and individual standing burnt trees. A wide strip of broken forest and burnt-out areas stretched for tens of kilometers from the southeast to the northwest to the Taseeva River and spread to its right bank, capturing the Angara interfluve. Local old-timers claimed that the fire with the windfall "flew" in 1908 from an unknown cause. The spruce taiga has been preserved only along the river valleys and their decays.""

The specified area is located approximately 500 km SW from the epicenter.

**The Tulun region forestfall**. Here is from (Vasil'ev, et al., 1981) (translated by A. Ol'khovatov):

"Pensioner Blinov G.T. in a letter dated 7 / IV-59, from the town of Georgievsk, Stavropol Territory, wrote (to Radio-committee, Moscow):
"The radio-broadcast brought me to the idea to inform everything that I know ( as a participant at that moment of what was visible, and after it became known to me due to circumstances, namely):
I don't remember the date, but I just remember that it was June 1908. At this point, I (being a worker) worked in Cheremkhovo, Irkutsk region. I worked with mountain trainees (students), worked under the roof, but not devoid of opportunity to see around, except for seeing over my head.
The day was clear, sunny, with just a slightly noticeable movement of air. The time was about 11 hours local time. Suddenly, a hurricane of the great strength pounced, accompanied by a strong hiss and whistle. Instantly the clouds of dust raised, the boards, sticks, clods of the earth flew. We failed to see anything above us because of dust. It took about 5-7 minutes, then everything got calm, as if nothing happened. Only in the air a turbid cloud of dust hung. "

Next, the author proceeds to the story that the commissioner of the Tulun District Executive Committee heard from the guide-driver. Regarding the hurricane in Cheremkhovo, there is no more word about it in the letter. In the north of the Tulun district, the driver "told the commissioner that in 1908, not without God's will, his brother died with his whole family and the entire household, from which literally nothing remained. An invisible force swept by, which cut down the top of the mountain, knocked down a lot of forest, and further in 20-30 versts in 3 places, a lot of forest burned out.""

The settlement of Cheremkhovo is situated about 860 km from the Tunguska epicenter at azimuth 174 deg.. Tulun is about 709 km from the epicenter at azimuth 188 degr., so the place of the given destructive event was rather far away from the epicenter.

**The Kazhma forestfall and some others located by satellite images.** Yu. D. Lavbin conducted expeditions in the mid 1990s in the Kazhma river region, where satellite images revealed possible forestfall. Lavbin delivered his report at the 1998 conference on the 1908 Tunguska event in Krasnoyarsk ( http://tunguska.tsc.ru/ru/science/conf/1998/6/9/ ). Here are some fragments of the report translated by A. Ol'khovatov:

"In the same year, 1996, I organized an expedition to this area, with the participation of a number of specialists. When approaching this area, from a helicopter, we clearly saw the directed fall of the taiga, the tops of the trees of which lie to the east – northeast. <...>

The forestfall, at its beginning, has a fan-shaped, finger-shaped character that passes further into a strip. The entire length of the forestfall is 45-50 km, with an average width of 5-7 kilometers. The total area of the forestfall is about 300 sq. km. The forestfall has a direction to the Southern swamp near Vanavara, i.e. to the Kulik's epicenter. The good preservation of the forestfall to this day is due to the fact that it almost completely extends along the tops of the hills and uplands, with a few exceptions in low-lying areas.

On the path of the forestfall, approximately in its central part, the top of the hill was "cut off", or rather destroyed, in which a scattering of large and small stones is observed on the eastern and north-eastern sides (examined from a helicopter). The height of the hill above sea level is 700 meters. At the moment, the top of the hill is flat, overgrown with young trees. <...>

During field work in the area of the epicenter and not far from it, trees were cut down, which showed that a space disaster in this area occurred in 1908, ... <...>

The analysis of the soil, samples of which were taken in this area, in different places, as well as the spherules found in it, showed the anomalous presence of many elements, some of which are truly cosmic. In particular, the percentage of Iridium has a value 3-4 orders of magnitude higher than its clark content in soils and rocks of the Earth. At the same time, a high content of such elements as: Germanium, Indium, Cobalt, Boron, Barium, Molybdenum, Manganese, Nickel, Lead, Copper, Magnesium, Zinc, Titanium, Sodium, Calcium, Phosphorus, etc., Iron 50 % (total) was found. <...>

The remoteness of this area from the Southern swamp near Vanavara is 250-270 km to the west, or rather to the southwest."

Lavbin also discovered other possible places of "spaceimpacts", and reported about them at the Krasnoyarsk conference devoted to 100th anniversary of Tunguska ( http://elib.sfu-kras.ru/handle/2311/8586 ).

Lavbin (he died in 2017) was not a scientist but a businessman, and he made some claims about discovering remnants of alien spaceship, so the author of this paper is cautious regarding his reports. Anyway as some respectable scientists took part in the Lavbin's expeditions, so it would be reasonable to conduct more research of the regions.

The Lavbin's reports became almost forgotten when a new article on this topic appeared in 2021. In (Gladysheva, and Yastrebov, 2021) unusual reliefs on the Earth's surface of Eastern Siberia were discussed which were detected on satellite images. Among them the "Kazhma formation" was mentioned. And also possibly some other forestfalls which were presented in this paper were mentioned too. The article (Gladysheva, and Yastrebov, 2021) ends with the following text:

"The sound and light effects of the Tunguska catastrophe were noted by local residents at distances of many hundreds of kilometers from the epicenter [2]. Considering the fact that the Tunguska cosmic body was a swarm of numerous fragments [3], it can be assumed that these Siberian fan reliefs are related to the Tunguska catastrophe.

Thus, at present, the reason for the formation of Siberian fan reliefs, as well as the time of their appearance on the Earth's surface, require further careful study."

**The Chuvar ("Western") forestfall.** On Fig.1 there are trial areas near x=40 km, y~0 km which belong to so called Chuvar forestfall. It was discovered in 1959 by KSE expedition (of course the forestfall was known to local residents a long time before). Here is how it was discovered from a book by KSE-members (Vasil'ev, et. al., 1960). The fragment says about a group of 3 KSE-researchers who explored a region to the west from the epicenter in the direction of the mountain marked on a map as 'height-593' (translated by A. Ol'khovatov):

"Soon the terrain began to rise gently in the direction to the north-west. There was no forestfall already a long ago - it was over five kilometers from the Khushma-river already. There was the old larch forest around, which densely overgrew the eastern margin of the foot of the height-593.

And suddenly the terrain landscape changed dramatically. The eastern slope was quite steep, and here, on this slope, the group again entered the forestfall zone, and even what the forestfall! Such a picture none of the three did see either on the Makikta-river, or to the south of the Hushma-river, or on the path of Kulik. The landscape seemed fantastic: huge trees, almost up to a meter across, were uprooted, thrown on top of each other, split like matches. All this, interlaced with red blocks of lichen-covered stones, seemed to be the trail of some devastating whirlwind, a tornado, a typhoon. At first the forestfall was completely disorderly; then (closer to the peak), trunks began to be located in a familiar way - the treetops in one direction, the roots in the other, but the direction of the fall of the trunks was directly opposite to that which is observed in the region investigated by Kulik: the treetops directed here to the east, and the roots to the west. The impression was created that the group encountered a new center of catastrophe, lying about thirty kilometers to the west of Kulik's huts."

In 1960 the Chuvar forestfall was investigated by a small group led by Leonid Innokent'evich Popov. He recalls that there were unusual signs of destruction on the top of the ridge. Most of the larches were bevelled without chips at a height of 0.5-0.7 meters from the ground. On the slope tops of trees were broken, and in the lowlands, the forest usually was not damaged. Popov recalled about his impressions of the Chuvar forestfall in his 2018 interview to the author of this paper ( [https://www.youtube.com/watch?v=QKn6GqxZxxs](https://www.youtube.com/watch?v=QKn6GqxZxxs) ). Popov (being an experienced taiga-traveler ) was puzzled that the trees on the top of the ridge were cut as if by a saw. Another surprising aspect was the direction of the fallen of trees, which was the opposite of what was expected. On Fig.9 here is a schematic drawing (to scale, with the exception of the Chuvar forestfall — its dimensions are increased for better viewing on the picture) of the Kulikovskii forestfall ( on the right), and the Chuvar forestfall ( on the left). The directions of the fallen trees (arrows) are also shown. Please pay attention that the scheme of directions of the fallen trees is simplified to better present the essence.

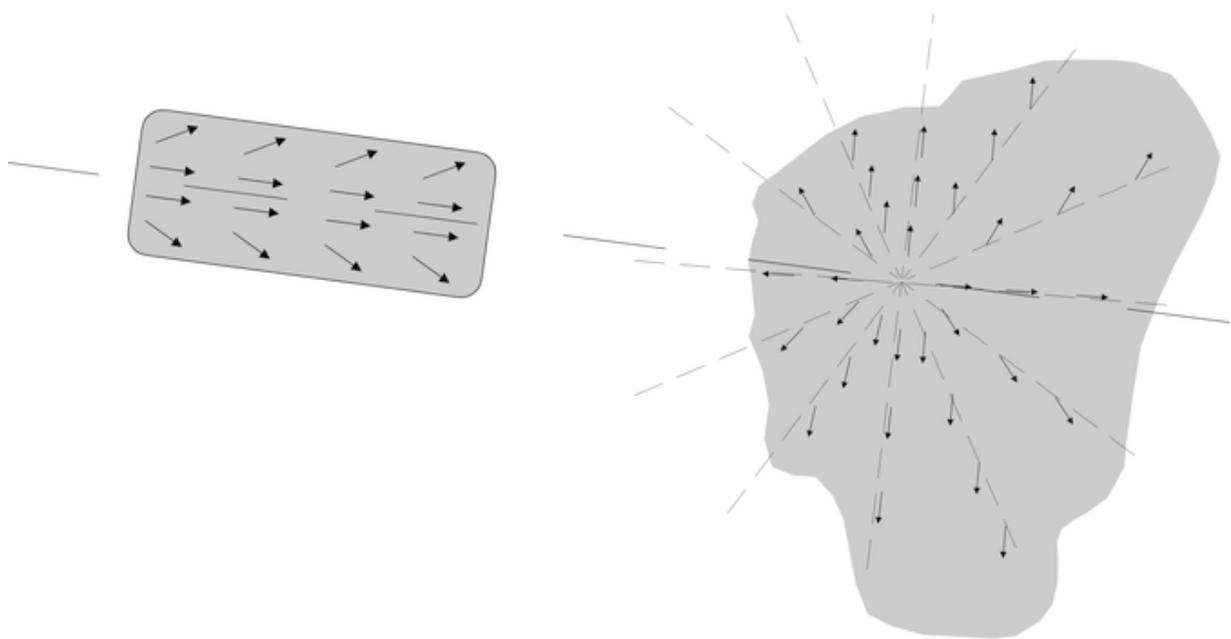

**Fig.9**

Popov also said in the interview that they searched for survived larch trees, cut them to count annual rings on the cuttings. The result was that the Chuvar forestfall was "about 1908" ( it was not possible to pinpoint better in field conditions). On Fig.10 there is a rare photo by Leonid Popov of trees (probably at the western foot of the Chuvar ridge). It is seen that just tops of the trees were broken.

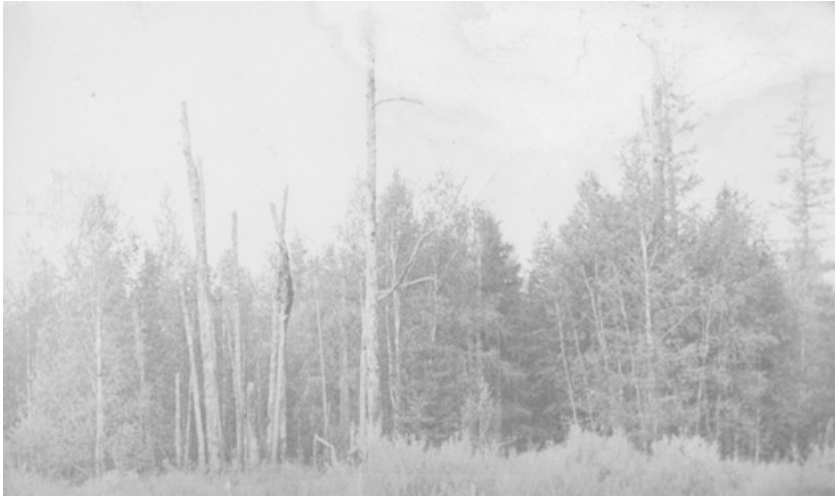

**Fig.10**

The photo was taken apparently in 1960 or 1961.

However in 1961 during a brief visit to the Chuvar forestfall another group of researchers discovered tree's damage which they assigned to the strong crown fire occurred at the end of the 19th century. So it was claimed that the Chuvar forestfall occurred as a result of a strong crown fire. However, how such fire could lead to the specific forestfall was not explained. Anyway this led to the termination of the Chuvar research for several years for concentrating research in the Kulikovskii forestfall area for alleged "meteoritic substance". And unfortunately a report by the Popov's group was not published.

In the mid 1960s Evenks accounts were collected which stated that burnt-out places were on the Chuvar ridge before and after the 1908 event, and that the Chuvar forestfall and the Kulikovskii forestfall occured in the same morning. Together with the failure to find the "remains of the Tunguska cosmic body" in the Kulikovskii forestfall area, this led to a renewed interest in the study of the Chuvar forestfall.

In 1971 the Chuvar forestfall was mapped. The map is shown on Fig.11 (see http://tunguska.tsc.ru/ru/archive/cae/1947/48-80/57/ for details)

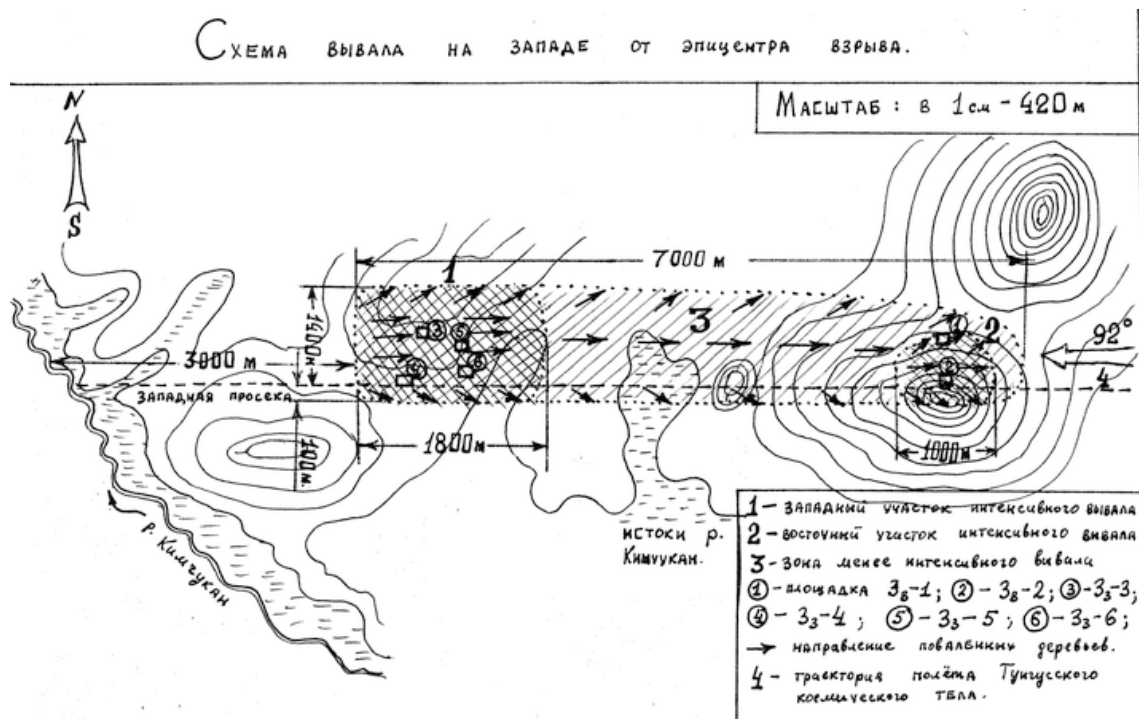

**Fig.11**

On Fig.11 marks 1,2 are areas of the most strong forestfall, mark 3 - less intensive.

According to the 1971 research this forestfall has a length of 7000 m, and a width of 1500 m. The boundaries of the forestfall are rather clear. According to the 1971 research: "It seemed that this small forestfall was "cut down" by an unknown force in the forest."

In 1981 Vasil'ev with coauthors summarized results as follows (Vasil'ev, et al., 1981) (translated by A. Ol'khovatov):

"Unlike the "eastern forestfall", the reality of which is doubtful, the "western forestfall" undoubtedly exists, and the question of its nature has not been completely removed from the agenda.
It was first discovered by the participants of the expedition in 1959. This is a fairly extensive (with an area of 30-40 km$^2$) section of the old windfall, located about 25 km to the west of the Kulik hut. Among the Evenks who witnessed the Tunguska disaster, there was talk that the "western forestfall" was formed at the same time as the fall of the forest in the area of Kulik's huts. In 1959-1971 it was repeatedly examined by route groups, who found that the age of this windfall is about the same as the fall of the forest caused by the Tunguska explosion, but its structure is completely different - the forestfall has a strip character, and the trees lie in the direction of the wind rose (mainly to the east). The old forest in this area has preserved distinct traces of a forest fire dating back to the beginning of the century. Currently, it is considered that the "western fall" is a forest burnt-out, the time of the appearance of which accidentally coincided with the Tunguska infall. At the same time, some participants of the expeditions (Yu. M. Emelyanov) believe that the origin of the "western forestfall" has a complex nature: the burnt-out in this area were formed several years before the disaster and have nothing to do with 1908, but the Tunguska explosion had some influence on this area (the appearance of tilted trees dating back to 1908).

In any case, it can be reliably stated that both in terms of configuration (strip structure) and in terms of orientation, the "western forestfall" differs significantly from the Kulikovskii one and most likely has no direct connection with the latter."

Thus, according to this interpretation, on the morning of June 30, 1908, two independent events occurred at a small distance from each other: the Tunguska spacebody infall that led to the Kulikovskii forestfall, and a wind that led to the specific Chuvar forestfall. However, apparently, realizing the weakness of such interpretation, the authors make a reservation with the words "most likely" and "the question of its nature has not been completely removed from the agenda".

Anyway the Chuvar forestfall exists and should be explained. By the way in 1991 traces of the 1908 fire-damage (on a tree) was discovered on the ridge Chuvar (Yashkov, and Krasavchikov, 2008; Yashkov, pers. comm, 2021).

## 4. Discussion

As follows from the above, the accuracy of the dating of these forestfalls differs significantly. The Chuvar forestfall and the Sulomay one have the best accuracy, while regarding the rest it can be said, at best, only "about June 30".

Unfortunately a little is known about the Sulomay forestfall, so let's discuss the

Chuvar one.

The situation with the Chuvar forestfall demonstrates the situation in which Tunguska research turned out to be in 1980s. Huge efforts to find the substance of "the Tunguska spacebody" did not bring unambiguous results. Moreover, numerous interviews with eyewitnesses led to unexpected results. Academician (of USSR Academy of Medical Sciences) Nikolai Vasil'ev, who was informal leader of the Tunguska research wrote (Vasil'ev, 1992) (a mistype is corrected):

> "Analysis of the catalog of statements by eyewitnesses to the disaster [11], the total number of which runs to a few hundred, reveals a fact that has not been clarified to date, namely that thunderlike sounds were heard not only during and after the flight of the bolide, but even before it. <...> It would hardly be realistic to explain them away as subjective errors, since claims of this kind are made over and over and independently of each other.
> <...>
> The second factor, a fairly odd factor, is related to the direction of motion of the body. Analysis of statements by witnesses who gathered along the hot tracks of the event [11] and in the 1920s and 1930s [25, 28] led the first investigators of the problem (L. A. Kulik, I. S. Astapovich, and E. L. Krinov) to the unanimous conclusion that the bolide traveled in the direction from south to north. However, analysis of the vector structure of the timber fall due to the shock wave of the Tunguska meteorite gives an azimuth of 114° [29, 30], and the field of burn damage even gives an azimuth of 95° [6-8], i.e., it indicates that the meteorite traveled from nearly east to west. It should be added that this direction also is confirmed by an analysis of the statements of eyewitnesses who lived at the time of the event in the upper reaches of the Lower Tunguska River (in the region of Preobrazhenka, Erbogachen, and Nepa). "

In 1994 Vasil'ev presented more details (Vasilyev, 1994) (TSB is Tunguska Spacebody):

> "The first investigators of the Tunguska meteorite (L.A.Kulik, E.L.Krinov, and I.S.Astapovich [1; 2; 3 ]) who analyzed comparatively fresh evidences of the flight of the TSB on the Angara river did not doubt that it had moved generally from the south to the north, though there were three versions of its trajectory (the southern one, proposed by L.A.Kulik, the south-eastern by

E.L.Krinov and the south-western by I.S.Astapovich). By the early 60-s it was Krinov's trajectory, namely 135º east of the true meridian, that was considered the most realistic.

Later however, as more information was accumulated on the vector structure of the fallen forest field [9; 17; 59], a "corridor" of axially symmetric deviations of the vectors of the forest falling from the dominating radial pattern was revealed, and this deviation was interpreted as the track of the ballistic wave. The direction of, the "corridor" which was initially estimated as 111º E from N (114º east of the true meridian) [17] was later found to be 95º E from N (99º east of the true meridian) [10], which roughly coincides with the axis of symmetry of the radiant burn area [19]. In this period of time, V. G.Konenkin [60] and later other investigators [61-63] questioned old residents of the area who had lived in the upper reaches of the Nizhnyaya (Lower) Tunguska in 1908 (where there was no questioning in the 20s and 30s). This resulted in the conclusion that TSB had been observed in the said area as well, the analysis of the data suggesting that the body moved from the ESE to the WNW, i.e. by the path coinciding with the projection of that of the TSB, as found on the basis of analysis of the vector picture of the fallen forest area. This coincidence caused revision of the notion of the TSB path, and since the year 1965 the ESE-WNW (in fact, even E-W) version has been accepted in literature. For some years it was assumed to be finally true.

A grave disadvantage of the calculations of TSB path before the mid-80s was that there were analyzed only some separate groups of eye-witnesses' accounts obtained by different researchers, in different periods of time, and not the whole body of evidence. Publication of the catalogue of eye-witnesses information [4] enabled analysis of the whole event. This was done in Ref. [56] and corroborated the considerations expressed earlier in Ref. [58] and also by I.S.Astapovich [64]. Two fundamental facts were established in particular:

1. The total combination of evidence given by "eye-witnesses of the Tunguska fall" contains in fact information on at least two (most likely more) large day-time bolides. It is important that the "images" of the "Angara" and the "Nizhnyaya Tunguska" bolides are quite different and everything seems to indicate that they belong to different objects.

2. The trajectory calculated on the basis of evidences of witnesses of the "Angara" phenomenon and corresponding most likely to its version proposed by E.L.Krinov [1] deviates considerably from that determined by analyzing of the vector structure of the forest fall area and the radiant burn area [9; 19]. Indeed, evidences of the Angara eye-witnesses, including the report of a district police officer, strongly suggest that the bolide flew "high in the sky", which is hardly consistent with the path 99º E of the true meridian. On the contrary, the data obtained on the Nizhnyaya Tunguska river, though agreeing with the configuration of the destruction area, are in contrast with the Angara observations.

An extra complication is that Nizhnyaya Tunguska data suggest virtually unambiguously that bolide's flight took place in the afternoon, unlike those of the Angara which refer to the early morning.

Attempts to resolve the conflict between the data face with considerable problems. If the Angara and Nizhnyaya Tunguska observations are due to different bolides, which is most probably so, then with which of them the destruction area originally explored by L.A.Kulik is associated? Judging by the destruction area configuration, the most probable candidate is the eastern (Nizhnyaya Tunguska) bolide. However none of TSB investigators doubts that the explosion at the distance of 70 km from Vanavara occurred in the early hours of the day, not past midday [56]. Moreover, there is no direct proof that the Nizhnyaya Tunguska bolide was observed in the year 1908, inasmuch as this event was not recorded in any official documents, unlike the Angara bolide.

Besides, even assuming the area of the leveled forest, discovered by L.A.Kulik, to be due to the Nizhnyaya Tunguska bolide, it remains unclear where the Angara bolide fell, then. Throughout the Tunguska "meteorite" study there was no doubt the latter had in fact exploded in the Vanavara region...

But if the forest leveling was caused by the Angara bolide, how does it fit the direction of the "corridor" impressed in the area of the fallen forest by the TSB ballistic wave?"

Vasil'ev underlines the following (Vasilyev, 1994 ):

"In the search of way out of this maze, more than one approach has been tried. Some researchers, preferring direct physical evidence, practically ignored eye-witnesses' testimonies as an unreliable subjective material. This approach could be agreed with to some extent, if it were a matter of a few inconsistent testimonies, not many hundreds of independent reports.

Besides - what is very important - the testimonies of the year 1908 include official documents of the time, whose authors were responsible to the authorities for their trustworthiness. For this reason, the eye-witnesses' reports should be regarded as a material equal to other data sets or at any rate not to be ignored, even if they do not conform to some speculative arguments."

This is only a small part of the problems with the interpretation of the Tunguska event, which were formed by the 1980s. An illustration of this can be a friendly cartoon by Valeriya Aleksandrovna Sapozhnikova (posted with her kind permission) on Fig.12. Nikolai Vasil'ev met mathematician Alyona Boyarkina on the winter Tomsk's street and discussion started on the situation that the Tunguska bolide can flew from 3 different directions.

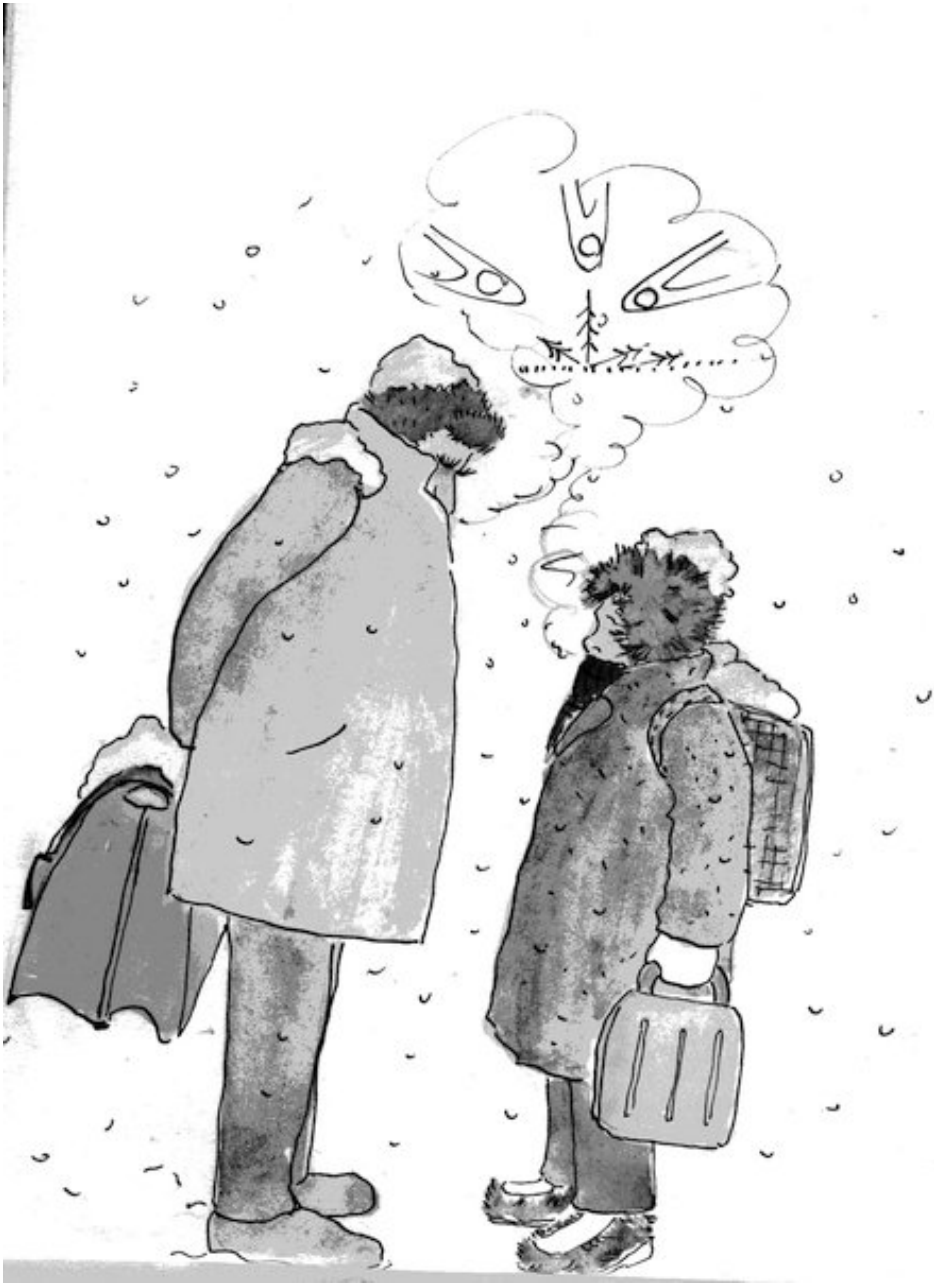

**Fig.12**

As it can be seen from the above, there is not even any certainty about "what flew, when, and where" on June 30. The addition of the Chuvar forestfall would complicate the already difficult situation many times over.

If all this not to consider just as "accidental coincidence", then the next interesting picture begins to look through. The Chuvar forestfall resembles a forestfall produced by a meteorological phenomenon called burst swath. This phenomenon was identified and classified only in the late 1970s by the prominent American meteorologist T. T. Fujita (Fujita, 1981).

In addition to the burst swath, Fujita also revealed another similar phenomenon, which he called a downburst. This phenomenon is generated by a jet of air, which after hitting the ground, spreads in all directions. So, the Kulikovskii forestfall resembles the one which is produced by a downburst. It is noteworthy that for the formation of both the Chuvar and the Kulikovskii forestfalls, the azimuths of the arrival of air jets are very close. Could the combination of these air-jets explain much weaker forestfall to the west (and NW) from the epicenter?

One more interesting aspect. Vasil'ev N.V. wrote (Vasil'ev, 1999) (translated by A. Ol'khovatov):

> "Further - today's vector field of the forestfall presents the field of the general, the main components of the prevailing vectors, while we all know well that the real distribution of the fallen trees on trial areas is polymodal. What to do with this polymodal, we have not come up with the last 30 years,..."

Probably one of the possible examples of the "polymodal" could be the next text by Krinov (Krinov, 1949) (translated by A. Ol'khovatov):

> "When viewing aerial photographs of the North-West, Western and South-Western sites, located at a distance of 2-4 km from the Southern swamp, that is, on the inner, very gentle slopes of the basin, there were places with a very powerful forestfall focused on the Southern swamp. However, in this site, a strip of almost total forestfall, focused on the North-West peatbog of the basin. This strip was detected by the Kulik back in 1927 and its direction is shown on its map (see fig. 14). Apparently, this collapse of the forest in the strip-form, not consistent with a common radial forestfall, was formed by a common strong hurricane, after the fall of the meteorite."

Could "this collapse of the forest in the strip-form" be caused by prolongation of the Chuvar air-jet? A food for thought…

Some more thoughts. William Corliss marked in his "Handbook of Unusual Natural Phenomena" sections on "Whirlwinds, explosive onset" and "Whirlwinds, fiery" (Corliss, 1977). In (Ol'khovatov, 2020b) there are several examples of whirlwinds with explosive onset occurred in the Tunguska event region. Also some regional weather info about June 30, 1908 is given in there, and some info on global geophysical situation is in (Ol'khovatov, 2003). More food for thoughts…

Anyway, discussion of possible interpretations of the 1908 Tunguska event is beyond the scope of this paper.

To once again demonstrate the complexity of the 1908 Tunguska event, here is an eyewitness testimony, which was included in the second (electronic) edition of the Tunguska witness accounts in 2018 (http://tunguska.tsc.ru/ru/science/1/eyewitness/):

> "From the State Archive of the Tomsk region (Cit. according to Doc No. 68)
> 3. Article by local historian E.I. Vladimirov "Kostromina zaimka"
> This area is located within the Rybinsk district of the Krasnoyarsk Territory, in 4 km south of 188 km of the Moscow road, 15 km from the station Solyanka, 25 km from the junction Filimonovo.
>
> After the appearance of my notes in the newspaper about the search for the Taseevsky and Tunguska meteorites, a resident of the village of Solyanka in the Rybinsk district of the Krasnoyarsk Territory Ivan Vasilyevich Gavrilyuk in July 1975 told me the following. His 86-year-old mother Praskovya Andreevna Korchak, who was born in 1889 on Kostroma Zaimka, often recalls the summer fall of a meteorite in 1908. She, on June 17, 1908, with her brothers Pyotr, 15 years old, Ivan, 12 years old, the younger Pyotr, 10 years old, came to the mowing that day early in the morning. Suddenly, with a clear sky, calm weather, deafening thunder rang out, fire flashed, dust rose, the Earth shook. The terrified children ran home past a strip of withered flax. Father Andrei Konstantinovich Korchak came to the scene of the incident. He saw that the flax plot had withered, as if scorched or frostbitten. There are fresh pits with pieces of stones nearby. Near the larch tree, at the fork of which there was an eagle's nest, the eaglets were scattered, lying on the Ground…
> When I heard about this, I remembered that in Adrianov's message it was said that a strong hum, a bright light, from the Filimonovo crossing, the peasants of the surrounding villages watched for 20-40 versts. Zaimka Kostroma is located 15 km from Solyanka and 25 km from Filimonovo to the southwest. I hurried to see Praskovya Andreevna. She confirmed what her son had said and, despite the illness of her legs, went with us to Kostromina Zaimka, indicated the place where the incident occurred on June 17, 1908."

Interestingly that the date is given in the Julian calendar, which hints that the witness remembers the original date. But if to propose that she could confuse with the exact date, then it means that peculiar events took place in the region during that times...

The author only wants to quote the final statement from the 1992 Vasil'ev article (Vasil'ev, 1992):

> "Since no final answer has been found to the question of the nature of the Tunguska phenomenon and since it must be acknowledged that many years of attempts to interpret it within the framework of the classical paradigm have not yet brought any decisive success, it seems worthwhile to examine and check alternative explanations."

And there is no "decisive success" within the framework of the classical paradigm 3 decades later…

## 5. Conclusion

There are some peculiarities of the Kulikovskii forestfall which any proposed interpretation of the Tunguska event should explain. Also there are arguments that the Tunguska event on June 30, 1908 was associated ( besides well-known Kulikovskii forestfall) with several smaller forestfalls. The level of argumentation is about the same as in the case of the Kulikovskii forestfall - Evenks accounts. There are also accounts (of local residents in general) pointing to several more small forestfalls, which dates can be pinpointed only less accurately, i.e. "about June 30", and so on. But as they were much smaller than the Kulikovskii one, then a little attention was paid to them.

Even if we take into account only the forestfalls with accurate dating, this already demonstrates the complex nature of the 1908 Tunguska event. Any interpretation of the Tunguska event should explain the facts.

**ACKNOWLEDGEMENTS**

The author wants to thank the many people who helped him to work on this paper,

and special gratitude to his mother - Ol'khovatova Olga Leonidovna (unfortunately she didn't live long enough to see this paper published...), without her moral and other diverse support this paper would hardly have been written.